\documentclass[preprint,  showkeys]{revtex4}
\usepackage{hyperref}
\usepackage{subfigure}
\usepackage{graphicx}
\usepackage{amsmath,amssymb}

\begin{document}

\title{Swampland Conjecture and Inflation Model from Brane Perspective}

\author{J. Sadeghi${}^{a}$}
\email{pouriya@ipm.ir}
\author{B. Pourhassan${}^{b}$}
\email{b.pourhassan@du.ac.ir}
\author{S. Noori Gashti${}^{a}$}
\email{saeed.noorigashti@stu.umz.ac.ir}
\author{S. Upadhyay${}^{c,d,e}$}
\email{sudhakerupadhyay@gmail.com}

\affiliation{${}^{a}$Department of Physics, Faculty of Basic Sciences, University of Mazandaran, P. O. Box 47416-95447, Babolsar, Iran.}
\affiliation{${}^{b}$School of Physics, Damghan University, P. O. Box 3671641167, Damghan, Iran.}
\affiliation{${}^{c}$Department of Physics, K.L.S. College,  Nawada, Bihar  805110, India.}
\affiliation{${}^{d}$Department of Physics, Magadh University, Bodh Gaya, Bihar 824234, India}
\affiliation{${}^{e}$Inter-University Centre for Astronomy and Astrophysics (IUCAA), Pune, Maharashtra 411007, India.}

\begin{abstract}
Over the past few decades, inflation models have been studied by researchers from different perspectives and conditions in order to introduce a model for the expanding universe. In this paper, we introduce a modified $f(R)$ gravitational model as ($R+\gamma R^{p}$) in order to  examine a new condition for inflation models. Given that our studies are related to a modified  $f(R)$ gravitational model on the brane, therefore we will encounter modified cosmological parameters. So, we first introduce these modified cosmological parameters such as spectral index, a number of e-folds and etc. Then, we apply these conditions to our modified $f(R)$ gravitational model in order to adapt to the swampland criteria. Finally, we determine the range of each of these parameters by plotting some figures and with respect to observable data such as Planck 2018.
\end{abstract}

\maketitle

\section{Introduction}

Recently, researchers have been using the weak gravity conjecture to study the inflation models as well as other cosmological implications. The weak gravity conjecture in the theories coupled with gravity   actually  describes  an important issue that gravity should be the weakest force. In fact, weak gravity conjecture is an important  tool to distinguish between two sectors of the effective low-energy theories: landscape (which are compatible with quantum gravity) and the swampland (which are incompatible with quantum gravity) \cite{1}. At the low energy level, the swampland is much wider than the landscape.  The landscape can actually be overlooked in front of the swampland at low energy levels. In that case, an important question is that what kinds of effective field theories are consistent with quantum gravity. Some of examples can be seen in Refs. \cite{R1,R2,R3}. The swampland has two conditions such as distance and dS conjecture which are examined in order to study inflation models according to specific conditions. For more information about weak gravity conjecture, landscape and swampland, one can visit  Refs.\cite{1,2,3,4,5,6,7,8,9,10,11,12,13,14,15}. Usually, the quantum gravitational corrections ignored in the eternal inflation studies and only the large-field models of inflation are considered. In order to consider the quantum gravitational corrections one should take in to account the small-field models for example via swampland conjecture. Of course, we need to mention a few very important points.
In general, swampland conjectures have not yet been widely accepted in the literature or it can not be considered as a complete theory, but by expressing conjectures and corrections to seek to resolve a series of ambiguities or in a way a proof for string theory.
 Although it has valuable answers in theories such as hot inflation and a few other examples, it still faces many challenges and many problems this theory.
However, it is possible in the future and with the further expansion of this theory and perhaps more observations that will have a good consistency theory about the many issues. But it is still a toddler that is growing and modifying and being used in various cosmological structures such as inflation, black hole physics, and dark energy. The second point; the quantum gravitational corrections are negligible in small-field models and very strong (potentially dominant) in large-field models. That is why the effective field theories associated with large-field models are not always reliable, while those corresponding to small-field models. Of course, in connection with the eternal inflation mentioned above, some points can also be added in this way, although the initial conjectures of the swampland are indicated by Matsui and Takahashi and Dimopoulos, which is generally incompatible with eternal inflation. But more recently William H. Kinney in his recent article (Eternal Inflation and the Refined Swampland Conjecture), has shown that the recently refined swampland conjecture, which somehow applies weaker criteria to the potential of the scale field in inflation, is somewhat consistent with eternal inflation \cite{2R1, 2R2, 2R3}. In fact, weak gravity conjecture, landscape  and swampland  are used to study inflation models and low-energy gravity theories in order to adapt or not to quantum gravity. The use of swampland criteria to study various inflation models in different theories of gravity has been studied by many researchers \cite{16,17,18,19,20,20-2}.

In the last few years, many inflation models have been evaluated from different aspects, such as the condition of slow-roll, constant-roll,  etc.   \cite{21,22,23,24,25,26, 26p}. To be precise, slow-roll models  depict  an early phase of
 de Sitter evolution of the Universe.
In order  to have  inflation  in effective low energy theories, inflationary models must be compatible to a UV complete field theory. As we know, the Swampland is
a set of  consistent effective field theories
which cannot be completed into quantum gravity in the UV.  Now,  in order to have particular compatibility with quantum gravity, or in other words to satisfy the quantum gravity,   an effective field theory   must satisfy the following  criteria
\cite{27,28,29,30,31}. The first swampland criterion (distance conjecture)
 which limits the range traversed by scalar
fields $\phi$ is
\begin{equation}\label{1}
\frac{\Delta\phi}{M_{pl}}<\Delta \sim \mathcal{O}(1),
\end{equation}
where $M_{pl}$ denotes to the  Planck mass.
Another swampland criterion (dS conjecture) is given by the gradient of the
potential $V$  for any scalar field satisfying the lower
bound  \cite{32,33}
\begin{equation}\label{2}
M_{pl}\frac{V'}{V}>c \sim \mathcal{O}(1),
\end{equation}
where $(c>0)$ and $'$ denotes the derivative with respect to $\phi$. According to the study of inflation models on the brane, the cosmological parameters must always have modifications, which we discuss in next section. However, the number of e-folds for single-field inflation is given by
\begin{equation}\label{3}
N=\frac{1}{M_{pl}^{2}}\int\frac{V}{V'}d\phi\approx\frac{\frac{\Delta\phi}{M_{pl}}}{M_{pl}(\frac{V'}{V})}.
\end{equation}
Recently, researchers have always used various methods to overcome the obstacles presented by these criteria to study these models and formalisms  which  can be seen in Refs. \cite{34,35}. In studying inflation models by using these criteria, according to previous works, we know that a number of inflation models are compatible with these criteria and some of the models are incompatible with them \cite{36}. Of course, there are models that are compatible with these criteria, but they are always associated with the problems in the initial conditions which has been extensively discussed in Refs. \cite{2002.02941} and \cite{1910.08837}. The objective of this paper is to improve the aforementioned  situation.

In section \ref{sec2}, we study the inflation model on the brane. In section \ref{sec3}, we introduce the modified  $f(R)$ gravitational model on brane. With respect to the concept proposed in section \ref{sec2}, we investigate the some cosmological parameters such as the $r$, $n_{s}$, $\alpha$  etc. Then we are plotting some figures to determine the concentrated areas of each parameter. Finally, in section \ref{sec4}, we explain the result of this paper.
\section{Inflation Model on Brane}\label{sec2}
In this section, we wish to express the modification of cosmological parameters according to  brane perspectives.
Inflation models have been studied from different perspectives. Given the models on brane as well as the concepts associated with braneworld, we know that our $4D$ world is a $3$-brane located in the higher dimensions of the bulk \cite{36,37,38,39}. Now, within the context of brane inflation, there exists series of modifications of cosmological parameters. The first relation in the brane scenario is the Friedman  equation, which  is modified to
\begin{equation}\label{4}
H^{2}=\frac{1}{3M_{pl}^2}\rho\left(1+\frac{\rho}{2\Lambda}\right).
\end{equation}
Here    $\Lambda$   is the three
dimensional brane tension which relates   Planck masses  as following:
\begin{equation}\label{5}
M_{4}=\sqrt{\frac{3}{4\pi}}\left(\frac{M_{5}^{2}}{\sqrt{\Lambda}}\right)M_{5},
\end{equation}
where $M_{4}$ is $4$-dimensional Planck mass scale and $M_{5}$ is $5$-dimensional Planck mass scale.

 Also the two  modified slow-roll parameters can be defined \cite{40}.  The first slow-roll parameter ($\epsilon$), which is a measure of the slope of the
potential, is given by
\begin{equation}\label{6}
\epsilon\equiv\frac{1}{2}\left(\frac{V'}{V}\right)^{2}\frac{1}{(1+\frac{V}{2\Lambda})^{2}}\left(1+\frac{V}{\Lambda}\right),
\end{equation}
and the second slow-roll parameter  ($\eta$) is given by
\begin{equation}\label{7}
\eta\equiv \left(\frac{V''}{V}\right)\left(\frac{1}{1+\frac{V}{2\Lambda}}\right).
\end{equation}

The number of e-folds ($N$) during inflation is as following \cite{40}
\begin{equation}\label{8}
N=- \frac{8 \pi}{M_4^2}\int_{\phi_{e}}^{\phi_{i}}\left(\frac{V}{V'}\right)\left(1+\frac{V}{2\Lambda}\right)d\phi.
\end{equation}
The power spectrum $(P_{R})$ is given by
\begin{equation}\label{9}
P_{R}=\frac{1}{12\pi^{2}}\frac{V^{3}}{V'^{2}}\left(1+\frac{V}{2\Lambda}\right)^{3},
\end{equation}
which is evaluated at horizon crossing.
Also, the scale-dependence of the perturbations is described by following scalar spectral index $(n_{s})$:
\begin{equation}\label{10}
n_{s}=1+2\eta-6\epsilon.
\end{equation}
In addition, the running spectral index, $\alpha$, has the following form:
\begin{equation}\label{11}
\alpha=\frac{dn_{s}}{d\ln k}=-\frac{dn_{s}}{dN}.
\end{equation}

Now, we apply the these concepts discussed in this section to the modified  $f(R)$ gravitational model. Then we determine the range of each parameter by plotting some figures and compare with observable data.

\section{Modified $f(R)$ gravitational model on the brane}\label{sec3}
Recently, people have described the characteristics of our universe by using the various inflation models. These modified inflation models play a very important role in describing the features of our universe, for example, in describing dark energy as well as cosmic acceleration. Different types of modified gravitational models are also used to describe quantum gravity. A number of these models are commonly used to describe neutron stars, as well as other types of them to study a variety of different cosmic models and gluons \cite{41,42,43,44,45,46,47,48}.   These inflation models are studied in different implications and conditions.
 Now, we want to examine a modified $f(R)$ gravitational model on the brane and by calculating the modified cosmological parameters, we study the compatibility of this model with swampland criteria. We consider a modified gravitational model with following expression of $f(R)$:
\begin{equation}\label{12}
f(R)=R+F(R)=R+\gamma R^{p},
\end{equation}
where $p$ is positive number (not necessarily integer). Also, $\gamma$ is a constant parameter, this coefficient solves the dimensional problem for such a model. Here we consider this value to be positive units for calculating the other important. Also, the result for this model with respect to $\gamma=1$ is very interesting that we explain in the following in detail. However, other values of $\gamma$ can be considered in an independent study.\\
First of all, we assume $V / \Lambda \gg 1$, so that the effect of brane is very significant. Hence, we should impose an upper bound to $\Lambda$. It should be noted that, if $V$ is decreasing during inflation, then assumptions such as $ V /\Lambda \gg 1 $ are satisfied at the end of inflation.\\
In order to investigate the potential of $f(R)$ gravitational model, we begin with the following Hilbert-Einstein  action:
\begin{equation}\label{13}
S=\int d^{4}x\sqrt{-g}\left[\frac{1}{2k^2}\widetilde{R}-\frac{1}{2}\widetilde{g}^{\mu\nu}\nabla_{\mu}\phi \nabla_{\nu}\phi - V(\phi)\right],
 \end{equation}
where $\widetilde{g}^{\mu\nu}= f'(R) {g}^{\mu\nu}$ and $\widetilde{R} =\varphi R=\exp\left(\sqrt{\frac{2}{3}}\phi R\right)$.
Now, we arrange the potential as
\begin{equation}\label{16}
V(\phi)=\frac{(\varphi-1)R(\phi)-F(R(\phi))}{2\varphi^{2}},
\end{equation}
and
we investigate the potential of $f(R)$ gravitational model, which is given by
\begin{equation}\label{17}
V(\phi)=\frac{1}{2}e^{-2\sqrt{\frac{2}{3}}\frac{\phi}{M_{pl}}}\left(-1+e^{\sqrt{\frac{2}{3}}\frac{\phi}{M_{pl}}}\right)^{\frac{p}{-1+p}}M_{pl}(p-1)p^{\frac{p}{1-p}}.
\end{equation}
Then,  according to equation (\ref{17}), we  have
\begin{equation}\label{18}
\frac{V'}{V}=-\frac{p(-\frac{\phi}{M_{pl}})^{-1+p}(-\frac{\phi}{M_{pl}})^{-p}\left(2^{\frac{3p}{2}}(-1+2p)(-\frac{\phi}{M_{pl}})^{-p}-3^{\frac{p}{2}}(-1+p)p!\right)}{M_{pl}(-1+p)\left(2^{\frac{3p}{2}}(-\frac{\phi}{M_{pl}})^{-p}-3^{\frac{p}{2}}p!\right)},
\end{equation}
where $\prime$ is the first derivative with respect to $\phi$.
The   second derivative of potential divided by potential leads to
\begin{equation}
\frac{V''}{V}=\frac{D(A+B)}{C},\label{19}
\end{equation}
where the explicit expression for $A$, $B$, $C$ and $D$ are, respectively,
\begin{eqnarray}
A&=&\left(M_{pl}p^{\frac{-2+p}{-1+p}}+2(1+p(-3+4p))\right)\left(-\frac{\phi}{M_{pl}}\right)^{2p},\\
 B&=&-2^{1+\frac{3p}{2}}*3^{\frac{p}{2}}(2+5(-1+p)p)\left(-\frac{\phi}{M_{pl}}\right)^{p}\Gamma(1+p)+2*3^{p}(-1+p)^{2}\Gamma^{2}(1+p),\\
 C&=&M_{pl}^{3}(-1+p)^{2}\left(2^{\frac{3p}{2}}\left(-\frac{\phi}{M_{pl}}\right)^{-p}-3^{\frac{p}{2}}p!\right)^{2},
 \\
 D&=&8^{p}P^{2+\frac{1}{p-1}}\left(-\frac{\phi}{M_{pl}}\right)^{-2}.
\end{eqnarray}
 In the following, we will assume $\frac{V}{\Lambda}\gg1$.  At the end of the section,   this assumption will turn out to be completely valid and correct.

According to equations (\ref{6}) and (\ref{7}), the modified slow-roll parameters
are calculated by
\begin{equation}\label{20}
\epsilon=\frac{2^{2-\frac{3p}{2}}3^{\frac{p}{2}}\Lambda p^{3+\frac{1}{-1+p}}p!(2^{3p}(-1+2p)^{2})(\frac{2^{\frac{3p}{2}}3^{-\frac{p}{2}}}{p!})^{-\frac{p}{-1+p}}(-\phi^{\frac{-2p^{2}-p+2}{p-1}})}{2^{3p}(-1+p)^{3}},
\end{equation}
\begin{equation}\label{21}
\eta=\frac{2^{2-\frac{3p}{2}}3^{\frac{p}{2}}\Lambda p^{3+\frac{2}{p-1}} 8^{p}(p^{\frac{p-2}{p-1}}+2(1+p(-3+4p))) p!(\frac{2^{\frac{3p}{2}}3^{-\frac{p}{2}}}{p!})^{\frac{p}{1-p}}(-\phi)^{\frac{-2P^{2}-p+2}{p-1}}}{2^{3p}(-1+p)^{3}}.
\end{equation}
Now the ratio of slow-roll parameters is
\begin{equation}\label{22}
\frac{\eta}{\epsilon}=\frac{p^{\frac{1}{p-1}}(2-6p+8p^{2}+p^{\frac{p-2}{p-1}})}{(1-2p)^{2}}.
\end{equation}
We know that the inflation ends at $\phi=\phi_{e}$  if one of the slow-roll parameters is of the order of one, that is, $\epsilon=1$ or $\eta=1$. With respect to above equation, it is obvious that $\eta>\epsilon$ for different values of $p$. Thus, we have
\begin{equation}\label{23}
(-\phi_{e})^{\frac{-2p^{2}-p+2}{p-1}}= \frac{(-1+p)^{3}2^{3p}}{2^{2-\frac{3p}{2}}3^{\frac{p}{2}}\Lambda p^{ \frac{3p-1}{p-1}} 8^{p}(p^{\frac{p-2}{p-1}}+2 +2p(-3+4p)  )p!(\frac{2^{\frac{3p}{2}}3^{-\frac{p}{2}}}{p!})^{\frac{p}{1-p}}(-\phi)^{\frac{-2P^{2}-p+2}{p-1}}}.
\end{equation}
Therefore, following the above computations and equation (\ref{8}), the number of e-folds  is calculated as (in unit of $\frac{8 \pi}{M_4^2}$)
\begin{equation}\label{24}
\begin{split}
N=&-\frac{2^{3p-2}3^{-p}(-1+p)^{3}p^{-2+\frac{1}{1-p}}(\frac{2^{\frac{3p}{2}}3^{-\frac{p}{2}}}{p!})^{\frac{1}{p-1}}(-\phi_{i})^{\frac{2P^{2}+p-2}{p-1}}}
{\Lambda(2-5p+4p^{3})(p!)^{2}}\\
&+\frac{2^{\frac{3p}{2}}3^{-\frac{p}{2}}p^{1+\frac{1}{1-p}+\frac{2}{p-1}}(p^{\frac{p-2}{p-1}}+2(1+p(-3+4p)))(\frac{2^{\frac{3p}{2}}3^{-\frac{p}{2}}}
{p!})^{\frac{p+1}{p-1}}}{(2-5p+4p^{3})(p!)}.
\end{split}
\end{equation}
If we ignore the second term of the number of e-folds, (the contribution of $\phi_{e}$), then (\ref{23}) becomes
\begin{equation}\label{25}
(-\phi_{i})^{\frac{2p^{2}+p-2}{p-1}}=\frac{\Lambda(2-5p+4p^{3})(p!)^{2}}{2^{3p-2}
3^{-p}(-1+p)^{3}p^{-2+\frac{1}{1-p}}(\frac{2^{\frac{3p}{2}}3^{-\frac{p}{2}}}{p!})^{\frac{1}{p-1}}}.
\end{equation}
Now, according to equations (\ref{25}),
the  $\epsilon$ in (\ref{20})  and $\eta$  in  (\ref{21}) take the following values, respectively:
\begin{eqnarray}
\epsilon &=& -\frac{2^{\frac{3p}{2}}+3^{-\frac{p}{2}}p^{1+\frac{1}{1-p}+\frac{1}{p-1}}(-1+2p)^{2}(\frac{2^{\frac{3p}{2}}*3^{-\frac{p}{2}}}{p!})^{\frac{1-p}{p-1}}}{(2-5p+4p^{3})N p!},\label{26}\\
\eta &=&-\frac{2^{\frac{3p}{2}}*3^{-\frac{p}{2}}p^{1+\frac{1}{1-p}+\frac{2}{p-1}}(p^{\frac{p-2}{p-1}}+2(1+p(-3+4p)))(\frac{2^{\frac{3p}{2}}*3^{-\frac{p}{2}}}{p!})^{\frac{p+1}{p-1}}}{(2-5p+4p^{3}) Np! }.\label{27}
\end{eqnarray}
 Now, corresponding to above expression of $\epsilon$ and $\eta$, we are interested to investigate the value of spectral index defined in (\ref{10}).  This leads to
 following spectral index:
\begin{equation}\label{28}
n_{s}=\frac{-26p^{2}+12p^{2+\frac{1}{p-1}}-16p^{3+\frac{1}{p-1}}-4p^{\frac{p}{p-1}}+p(6-5N)+2N+4p^{3}(6+N)}{(2-5p+4p^{3})N}.
\end{equation}
In order to check the validity of   assumption made for $\Lambda$, the very interesting and exciting point is that, in the above calculations, one can see the cosmological parameters are only a function of ($p$) and ($N$). They don't depend on the parameter $ \Lambda $. So we can say that the initial assumption of $\Lambda$ is completely correct and acceptable.

Now, according to equation   (\ref{28}), we obtain the expression for running spectral index (\ref{11}) as follows,
\begin{equation}\label{29}
\alpha=\frac{-4p^{\frac{p}{p-1}}-2p(-3+4p)(1+p(-3+2p^{\frac{1}{p-1}}))}{(2-5p+4p^{3})N^{2}}.
\end{equation}
Exploiting
expression  (\ref{25}),
the spectrum (\ref{9}) is calculated as
\begin{equation}
P_{R}=\frac{\mathcal{L}\cdot\mathcal{M}\cdot\mathcal{N}}{O},\label{30}
\end{equation}
where
\begin{eqnarray}
\mathcal{L}&=&3^{-1-2p}8^{-3+2p}(-1+p)^{6}p^{ \frac{-6p+2}{p-1}} \times\nonumber\\
&&\left(-\frac{\Lambda(2-5p+4p^{3})p!^{2}N}{2^{-2+3p}
3^{-p}(-1+p)^{3}p^{-2+\frac{1}{1-p}}(\frac{2^{\frac{3p}{2}}3^{-\frac{p}{2}}}{p!})^{\frac{1}{p-1}}}\right)^{\frac{(p-1)(4p+2)}{2p^{2}+p-2}}, \\
\mathcal{M}&=&\left(-1+ 2^{\frac{3p}{2}}3^{-\frac{p}{2}} \left(-\frac{\Lambda(2-5p+4p^{3})p!^{2}N}{2^{-2+3p}
3^{-p}(-1+p)^{3}p^{-2+\frac{1}{1-p}}(\frac{2^{\frac{3p}{2}}3^{-\frac{p}{2}}}{p!})^{\frac{1}{p-1}}p!}\right)^{\frac{p(p-1)}{2p^{2}+p-2}}   \right)^{\frac{4p}{p-1}},\\
\mathcal{N}&=& 2^{ {3p} } \left(-\frac{\Lambda(2-5p+4p^{3})p!^{2}N}{2^{-2+3p}3^{-p}(-1+p)^{3}p^{-2+\frac{1}{1-p}}(\frac{2^{\frac{3p}{2}}
3^{-\frac{p}{2}}}{p!})^{\frac{1}{p-1}}}\right)^{\frac{p(p-1)}{2p^{2}+p-2}} -3^{\frac{p}{2}}p!, \\
O&=&\Lambda^{3}\pi^{2}(p!)^{4}(2^{\frac{3p}{2}}(-1+2p) \left(-\frac{\Lambda(2-5p+4p^{3})p!^{2}N}{2^{-2+3p}*3^{-p}(-1+p)^{3}p^{ \frac{-1+p}{1-p}}(\frac{2^{\frac{3p}{2}}3^{-\frac{p}{2}}}{p!})^{\frac{1}{p-1}}}\right)^{\frac{p(p-1)}{2p^{2}+p-2}} \nonumber\\
&-&3^{\frac{p}{2}}(-1+p)p!)^{2}.
\end{eqnarray}
Also, the tensor-to-scalar ratio \cite{49,50} is obtained with respect to equation (\ref{26}) as
\begin{equation}\label{31}
r=24\epsilon =-24\left(\frac{2^{\frac{3p}{2}}+3^{-\frac{p}{2}}p^{1+\frac{1}{1-p}+\frac{1}{p-1}}(-1+2p)^{2}(\frac{2^{\frac{3p}{2}}*3^{-\frac{p}{2}}}{p!})^{\frac{1-p}{p-1}}}{(2-5p+4p^{3})N p!}\right).
\end{equation}
After calculating the modified cosmological parameters, we  plot some figures
of each parameter in accordance to the computational results and observable data.
The plotted figures and their explanations are as following.

\begin{figure}[h!]
 \begin{center}
  {
 \includegraphics[height=6cm,width=6cm]{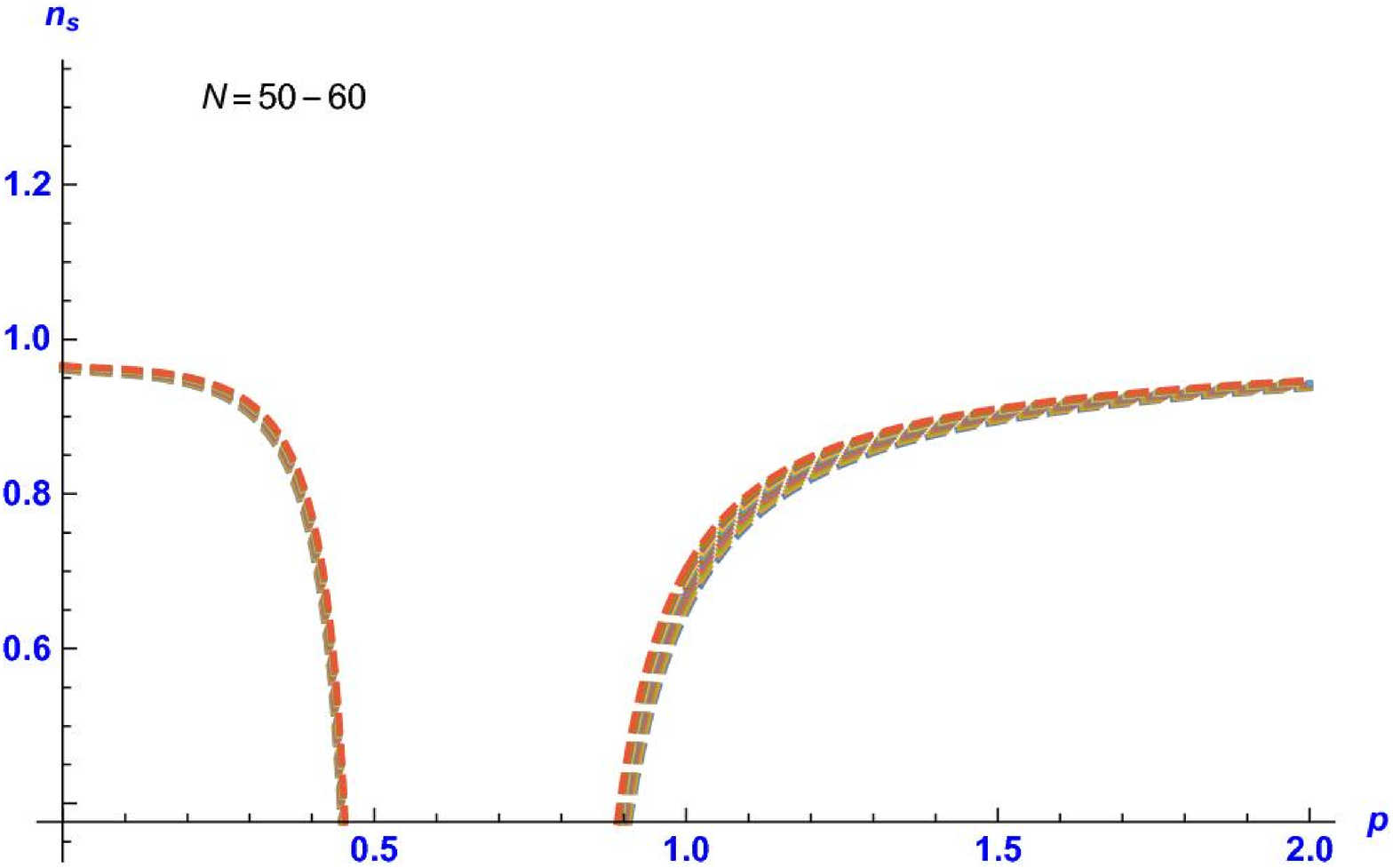}
 \label{1a}}
 \caption{The spectral index $n_{s}$ in term of $p$ for number of e-folds
  ($N=50-60$).}
 \label{fig1}
 \end{center}
 \end{figure}

\begin{figure}[h!]
 \begin{center}
 {
 \includegraphics[height=6cm,width=6cm]{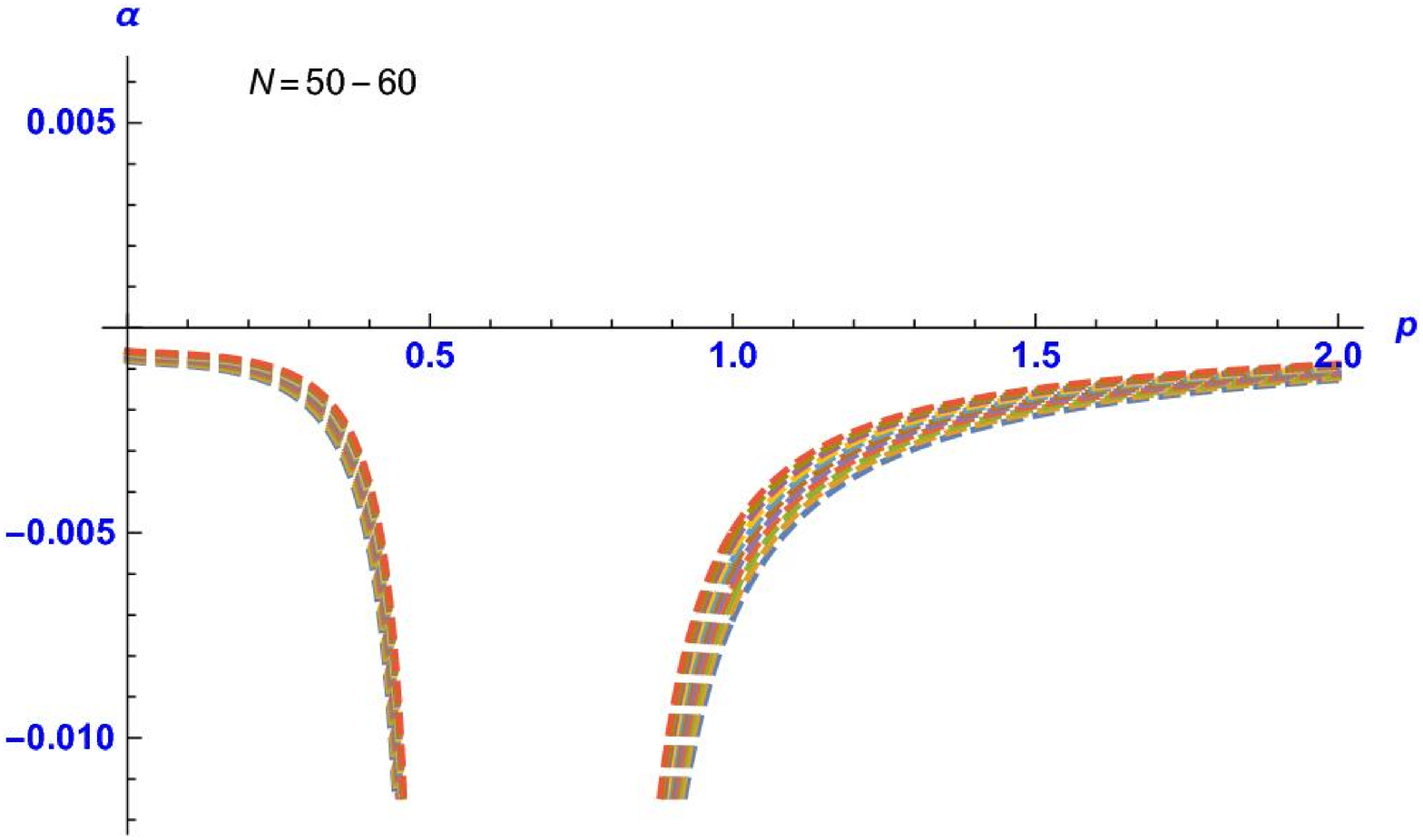}
 \label{2a}}
 \caption{The running spectral index $\alpha$ in term of $p$ for number of e-folds ($N=50-60$).}
 \label{fig2}
 \end{center}
 \end{figure}

\begin{figure}[h!]
 \begin{center}
  \subfigure[] {
 \includegraphics[height=5cm,width=5cm]{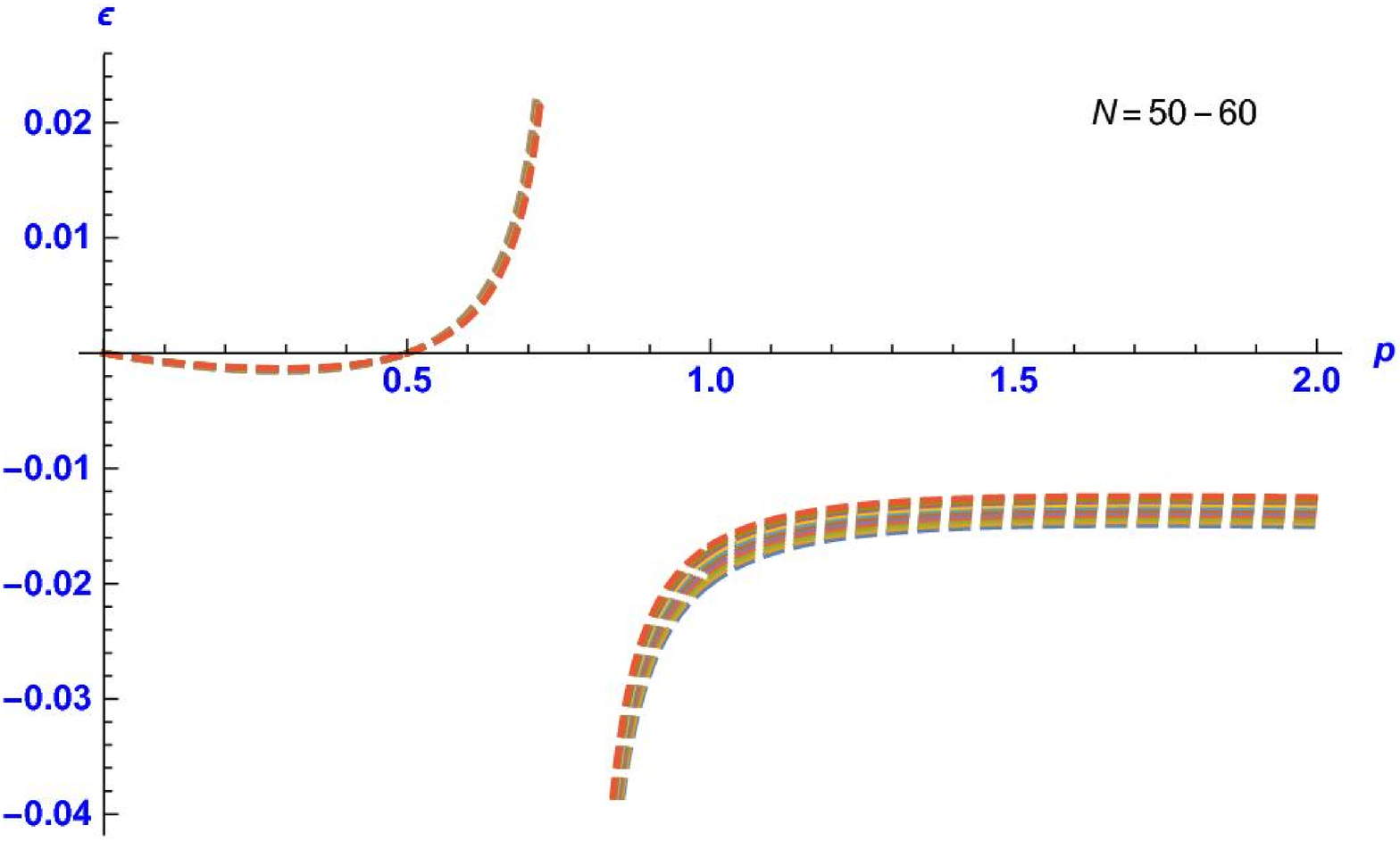}
 \label{3a}}
  \subfigure[]{
 \includegraphics[height=5cm,width=5cm]{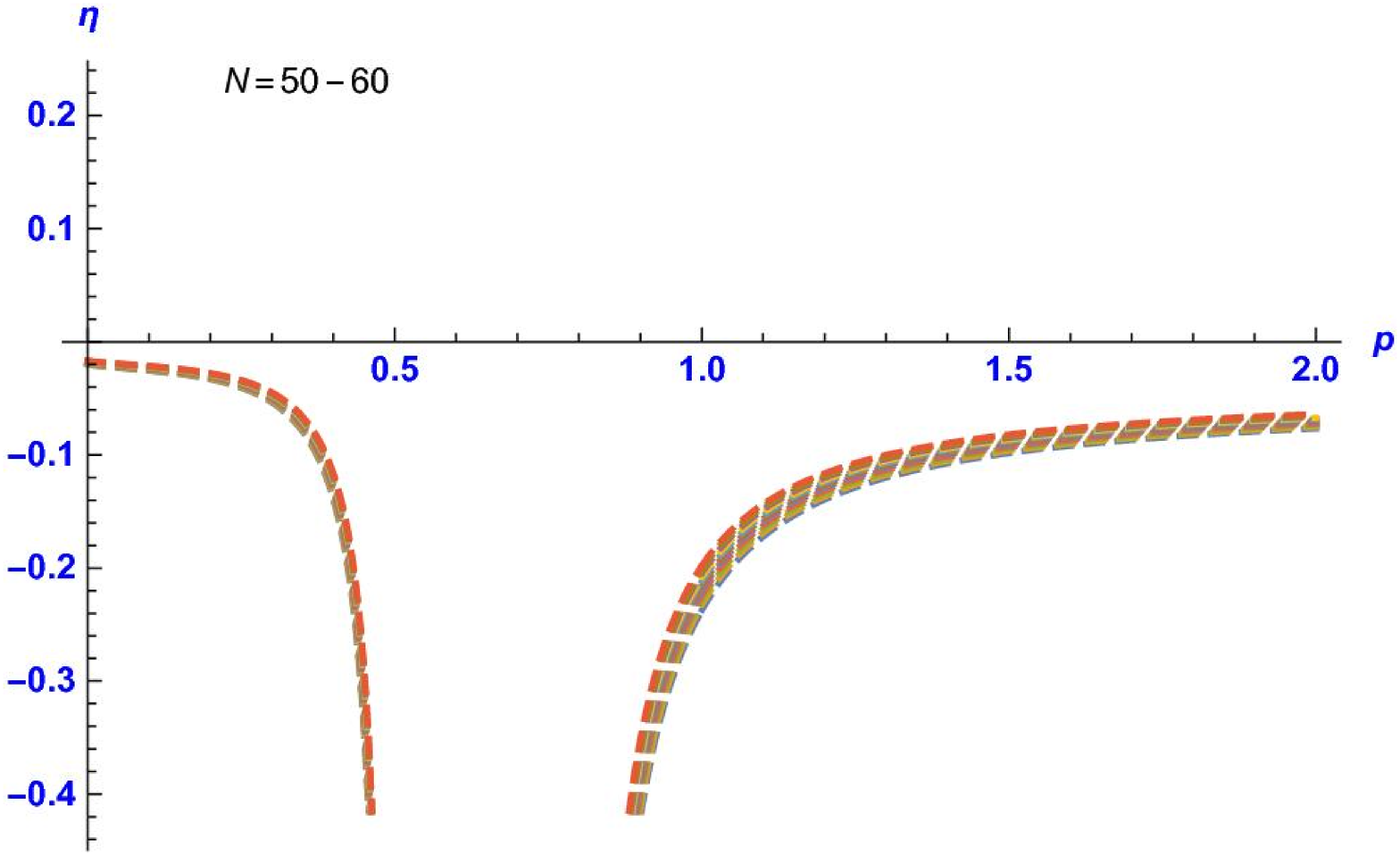}
 \label{3b}}
  \caption{\small{The slow-roll parameters $\epsilon$ in (a) and $\eta$ in (b) for  $p$  and $N=50-60$. }}
 \label{fig3}
 \end{center}
 \end{figure}

The figures \ref{fig1}, \ref{fig2} and \ref{fig3} depict  the range associated with the scalar spectrum index, the running spectrum index and the slow roll parameters, respectively,  for different values of $p$ and number of e-folds ($N$).
The range of each parameter  specified in these plots   is consistent with the observable data \cite{51}.
In Fig. \ref{4}, we plot $\eta$ with respect to $\epsilon$ for $N=60$.

\begin{figure}[h!]
 \begin{center}
 {
 \includegraphics[height=6cm,width=5cm]{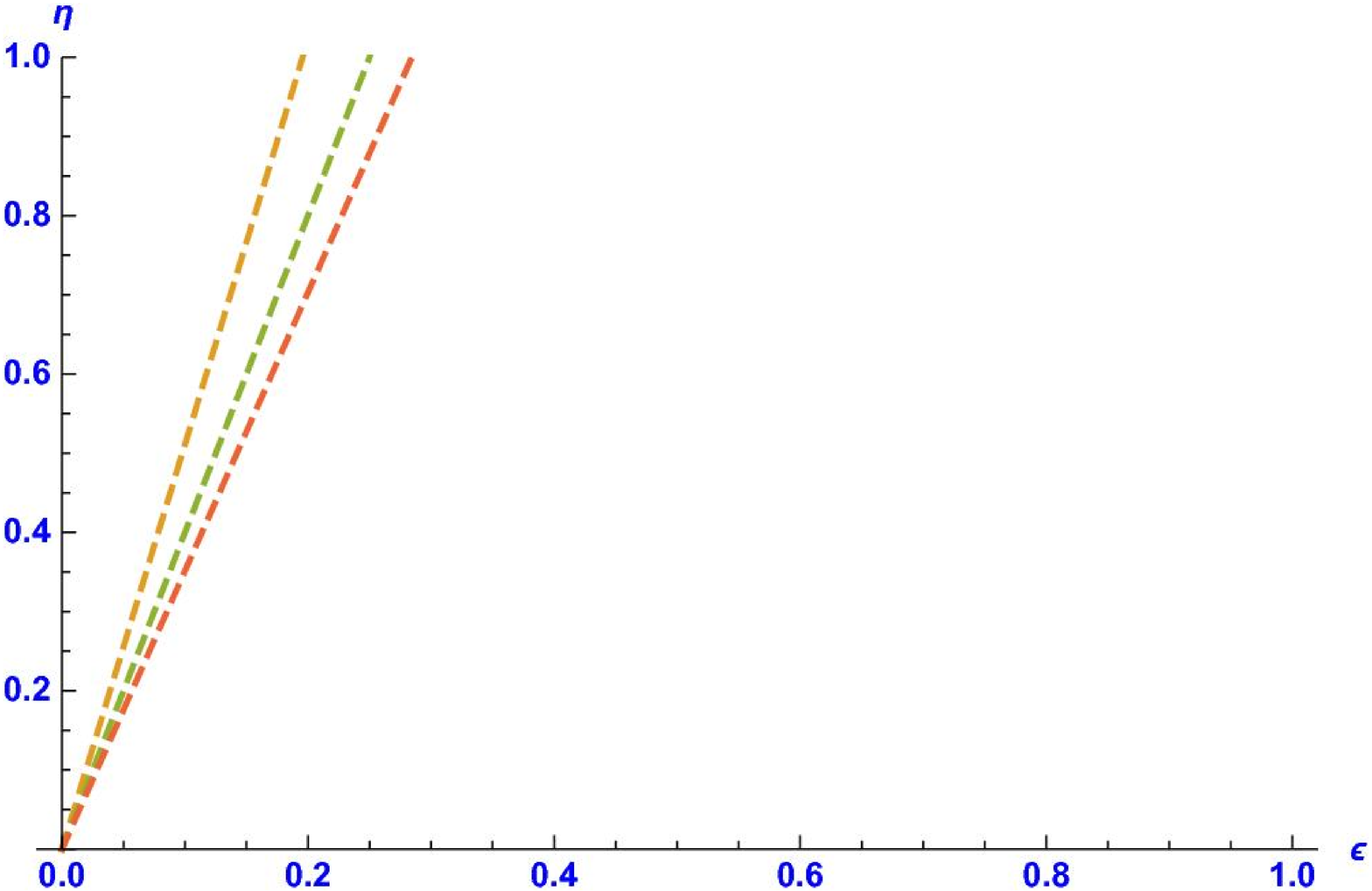}
 }
   \caption{\small{The plot of slow-roll parameter  $\eta$ in terms of  $\epsilon$  for $N=60$ and different values of $p$ as brown ($p = 3$), green ($p =4$), and red ($p= 5$). }}
 \label{fig4}
 \end{center}
 \end{figure}
 \begin{figure}[h!]
 \begin{center}
   \subfigure[]{
 \includegraphics[height=5cm,width=5cm]{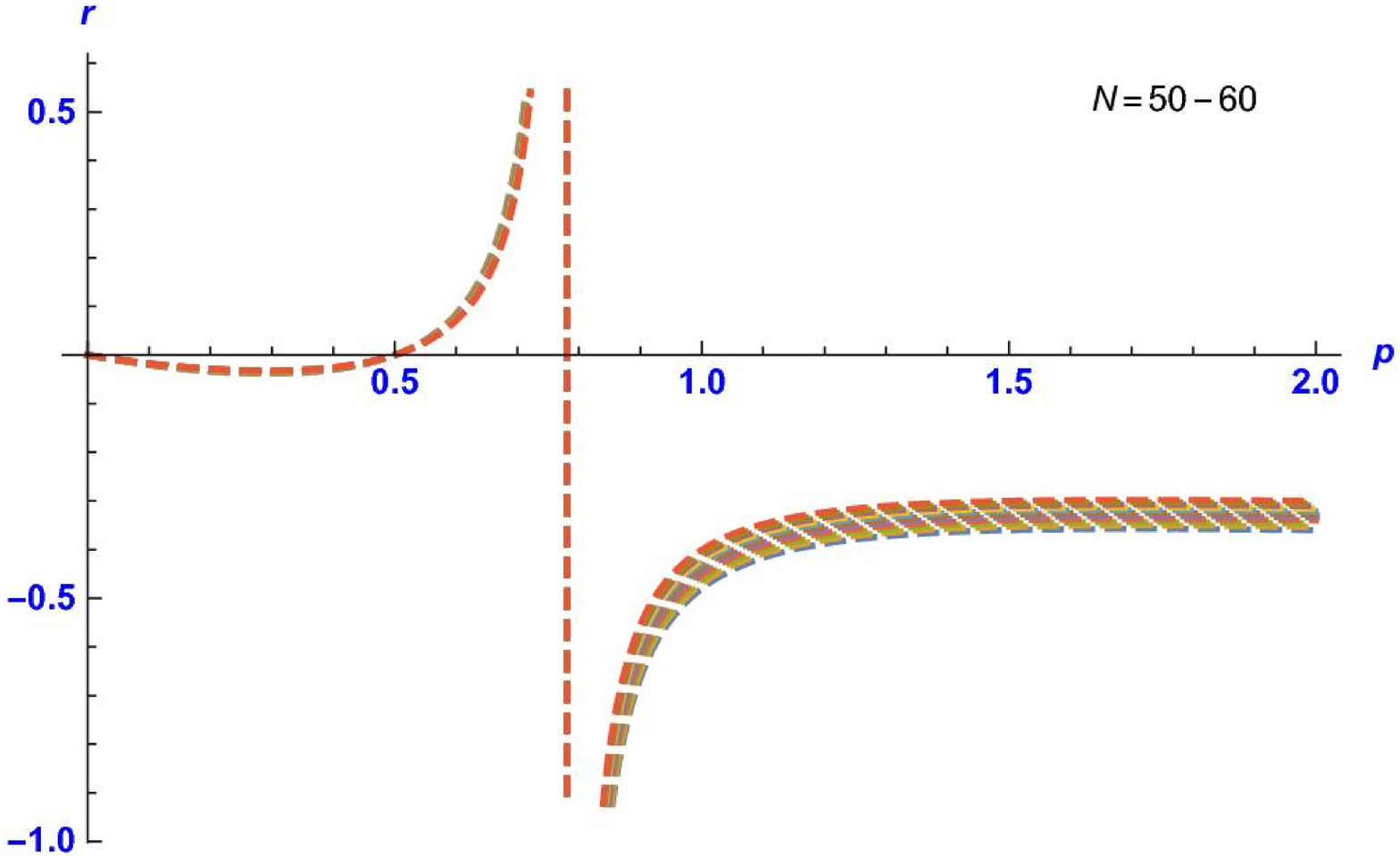}
 \label{5a}}
   \subfigure[]{
 \includegraphics[height=6cm,width=5cm]{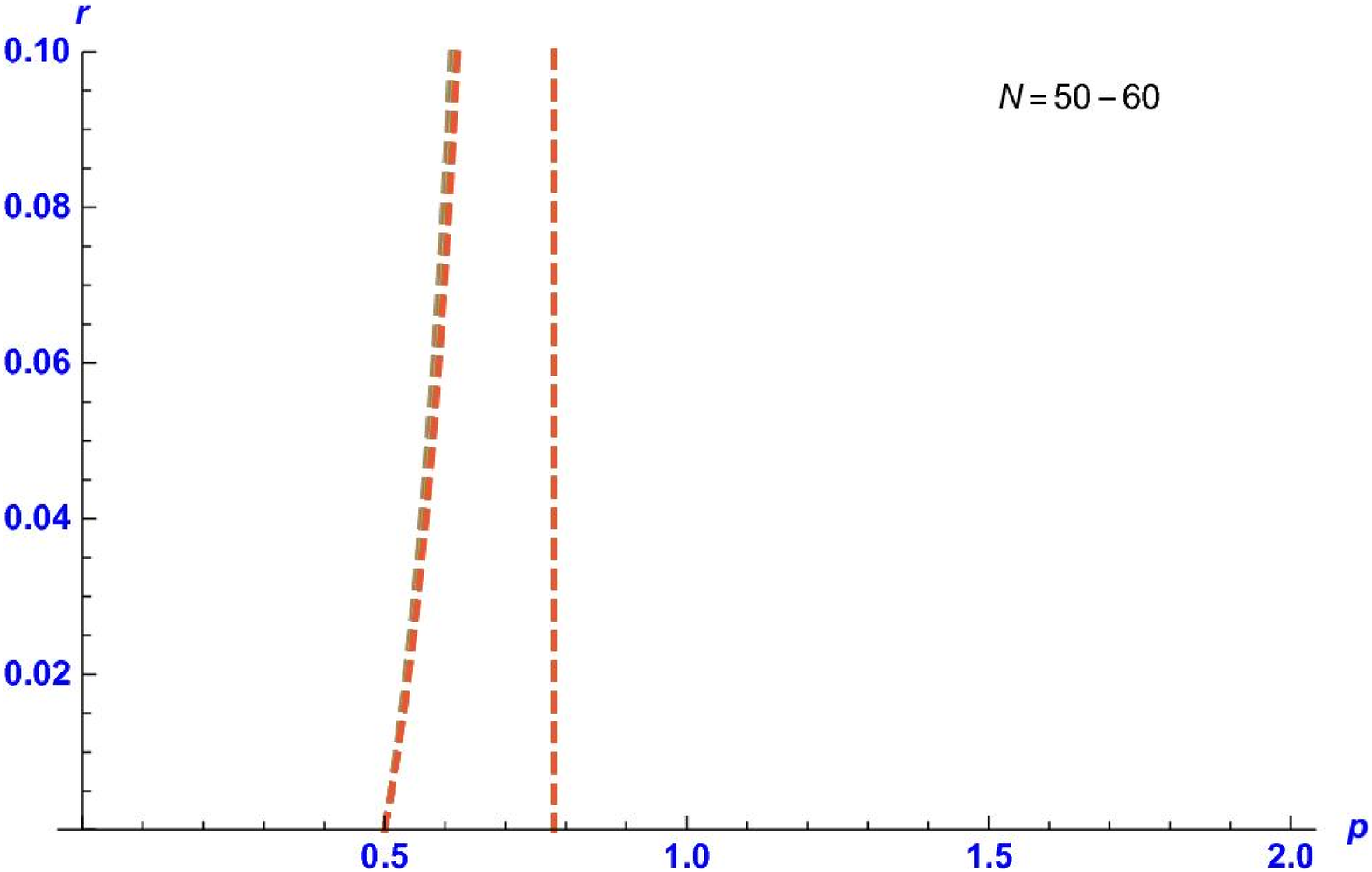}
 \label{5b}}
  \caption{\small{The plot of  $r$ in terms of $p$  in  (a) for $N=50-60$ and  (b) is close up of (a).}}
 \label{fig5}
 \end{center}
 \end{figure}

\begin{figure}[h!]
 \begin{center}
    \subfigure[]{
 \includegraphics[height=5cm,width=5cm]{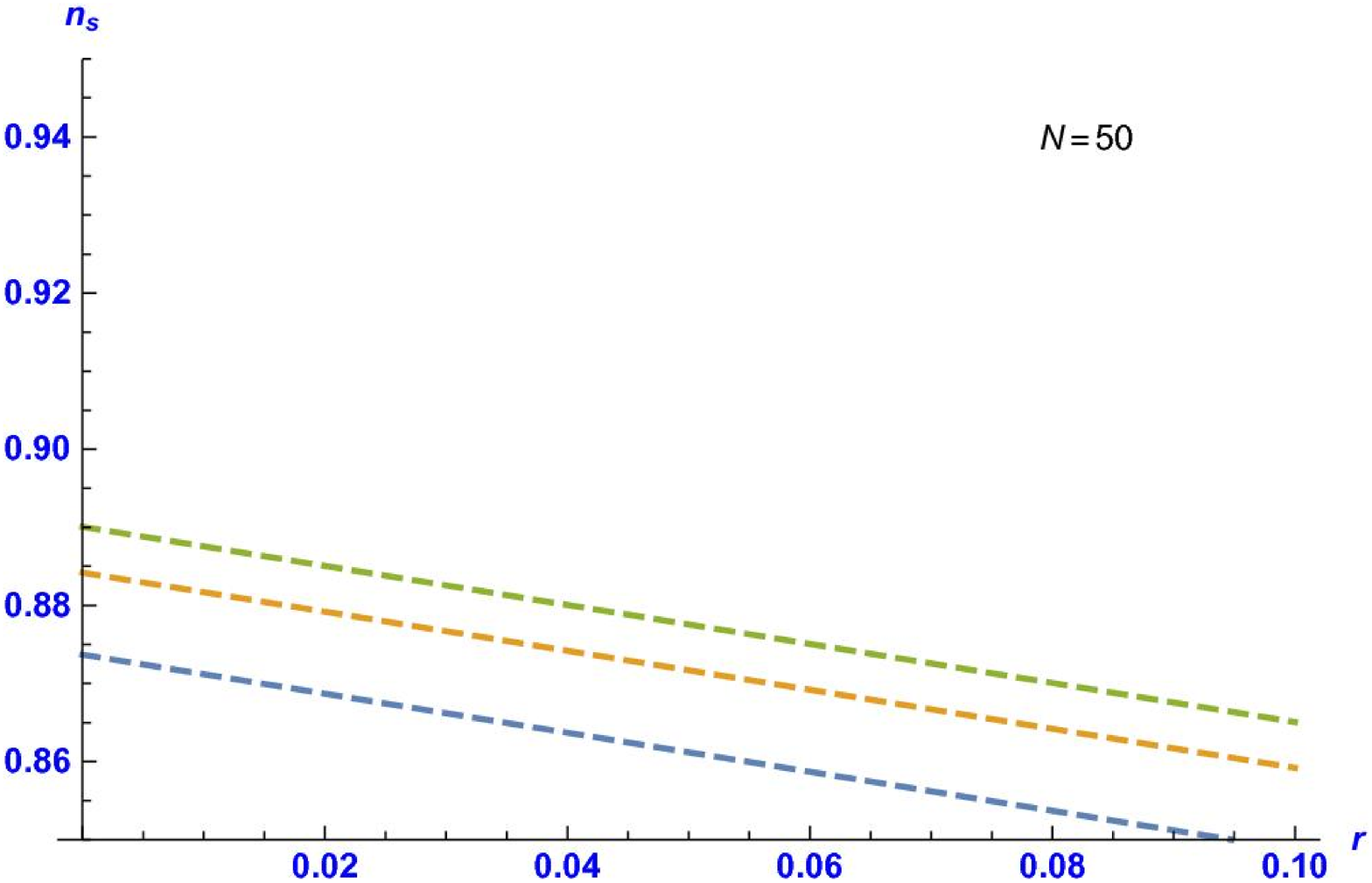}
 \label{6a}}  \subfigure[]{
 \includegraphics[height=5cm,width=5cm]{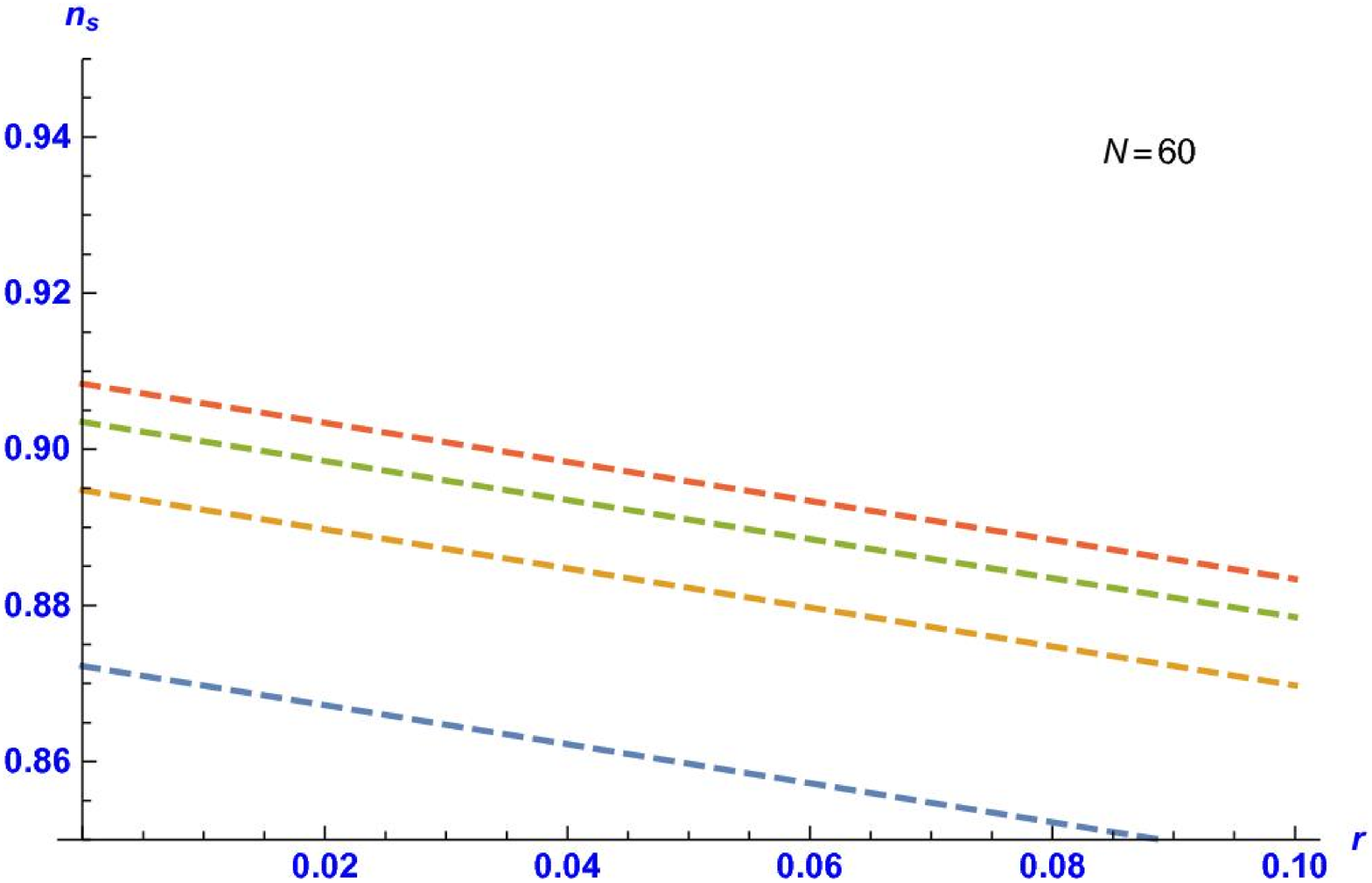}
 \label{6b}}
   \subfigure[]{
 \includegraphics[height=5cm,width=5cm]{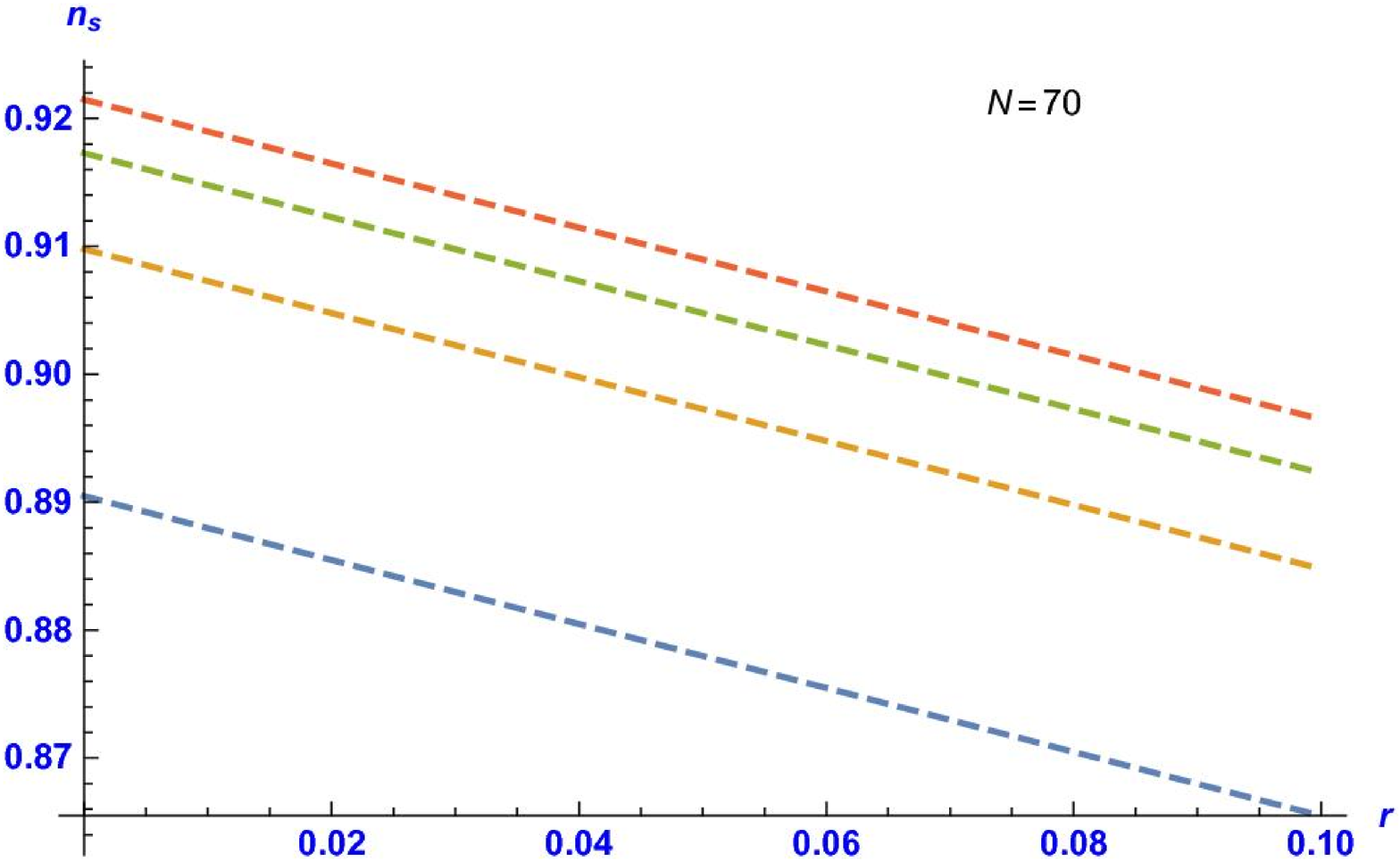}
 \label{6c}}
  \caption{\small{The plots of $n_{s}$ in terms of $r$  for $N=50$,  $N=60$ and $N=70$ and different values of $p$ as blue ($p=2$), brown ($p =3$), green ($p =4$) and red ($p=5$). }}
 \label{fig6}
 \end{center}
 \end{figure}
We plot  some figures to investigate each of these cosmological parameters, according to the different values of  $p$,  $N$   and observable data.  We plot the $r$ in term of  $p$ with respect to $N=50-60$ in Fig.  \ref{fig5}. The validity of this parameter is determined according to Planck's data which is $0.064$.  Also, we plot $n_{s}$ plan for $N=50$, $N=60$ and $N=70$ and different value of $p$  in figure \ref{fig6}. As shown in the figures, the values are closer to the observable data for number of e-folds ($N = 50, 60$).

Now,  if we consider the second term of the number of e-folds or the contribution of $\phi_{e}$ corresponding to (\ref{24}), then
\begin{eqnarray}\label{32}
 (-\phi)_{i}^{\frac{2p^{2}+p-2}{p-1}}&=& \frac{2^{2-3p}*3^{p}\Lambda p^{\frac{2p}{p-1}}}{(-1+p)^{3}}
 \left[2p(1+p(-3+4p))\right.\nonumber\\
&+&  \left. p^{\frac{1}{1-p}}(p^{2}+N(-2+5p-4p^{3}))\right](\frac{2^{\frac{3p}{2}}3^{-\frac{p}{2}}}{p!})^{\frac{1}{1-p}}p!^{2}.
\end{eqnarray}

Therefore, the modified slow-roll parameters given in the equations (\ref{26}) and (\ref{27}) convert, respectively, to
\begin{equation}\label{33}
\epsilon=\frac{(1-2p)^{2}p}{p^{2}-6p^{2+\frac{1}{p-1}}+8p^{3+\frac{1}{p-1}}=2p^{\frac{p}{p-1}}-2N+5pN-4p^{3}N},
\end{equation}
and
\begin{equation}\label{34}
\eta=\frac{p^{\frac{p-2}{p-1}}(2p^{\frac{2}{p-1}}+p^{\frac{p}{p-1}}+8p^{\frac{2p}{p-1}}-6p^{\frac{1+p}{p-1}}}{p^{2}-6p^{2+\frac{1}{p-1}}+2p^{\frac{p}{p-1}}-2N+5Np-4P^{3}N}.
\end{equation}
The  the spectral index (\ref{28})  with respect to the new condition becomes
\begin{equation}\label{35}
n_{s}=3+\frac{6(1-2p)^{2}p-2(2-5p+4p^{3})N}{-2p^{\frac{p}{p-1}}+2N+p(-5N+p(-1-2p^{\frac{1}{p-1}}(-3+4p)+4pN))},
\end{equation}
and the running spectral index (\ref{29})  with respect to the new condition  becomes
\begin{equation}\label{36}
\alpha=\frac{2(2-5p+4p^{3})(2p^{\frac{p}{p-1}}+p(-3+4p)(1+p(-3+2p^{\frac{1}{p-1}})))}{(p^{2}-6p^{2+\frac{1}{p-1}}+8p^{3+\frac{1}{p-1}}+2p^{\frac{p}{p-1}}-2N+5pN-4p^{3}N){2}}.
\end{equation}
Also, the tensor-to-scalar ratio takes the following form:
\begin{equation}\label{37}
r=24\left(\frac{(1-2p)^{2}p}{p^{2}-6p^{2+\frac{1}{p-1}}+8p^{3+\frac{1}{p-1}}+2p^{\frac{p}{p-1}}-2N+5pN-4p^{3}N}\right).
\end{equation}
As you can see from the above equations, by considering the contribution of $\phi_{e}$ from equation (\ref{28}) the cosmological parameters are only a function of two values of  $p$ and $N$ and independent of $\Lambda$. Actually, it is consistent with our initial assumption of $\Lambda$. According to the observable data and the above equations, we determine the range of each of these parameters from the figures. Then we compare the changes between these two parts, i.e. by considering the contribution of  $\phi_{e}$ and without it. In general, the final results will be very close to each other. Also, according to equations (\ref{9}) and (\ref{32}), we again calculate the values of $P_{R}$  given as
\begin{equation}\label{38}
\begin{split}
&\phi=(\frac{2^{2-3p}3^{p}\Lambda p^{\frac{2p}{p-1}}(2p(1+p(-3+4p))+p^{\frac{1}{p-1}}(p^{2}+N(-2+5p-4p^{3})))(\frac{2^{\frac{3p}{2}}
3^{-\frac{p}{2}}}{p!})^{\frac{1}{1-p}}(p!)^{2}}{(p-1)^{3}})^{\frac{p-1}{2P^{2}+p-2}}\\
&P_{R}=\frac{3^{-1-2p}8^{-3+2p}(-1+p)^{6}p^{-6-\frac{4}{p-1}}(\phi)^{2+4p}(-1+\frac{2^{\frac{3p}{2}}
3^{-\frac{p}{2}}(\phi)^{p}}{p!})^{\frac{4p}{p-1}}(2^{\frac{3p}{2}(\phi)^{p}-3^{\frac{p}{2}}p!})^{2}}{\Lambda^{3}\pi^{2}(p!)^{4}(2^{\frac{3p}{2}}(-1+2p)(\phi)^{p}-3^{\frac{p}{2}}(-1+p)p!)^{2}}
\end{split}
\end{equation}

Now, we plot the range of each cosmological parameter  specified by taking the contribution of $\phi_{e}$ into account  and  the observable data and measurement values.  With respect to the above description, the range of each cosmological parameter with regard to the new conditions in  plots will be been determined.
\begin{figure}[h!]
 \begin{center}
 {
 \includegraphics[height=5cm,width=5cm]{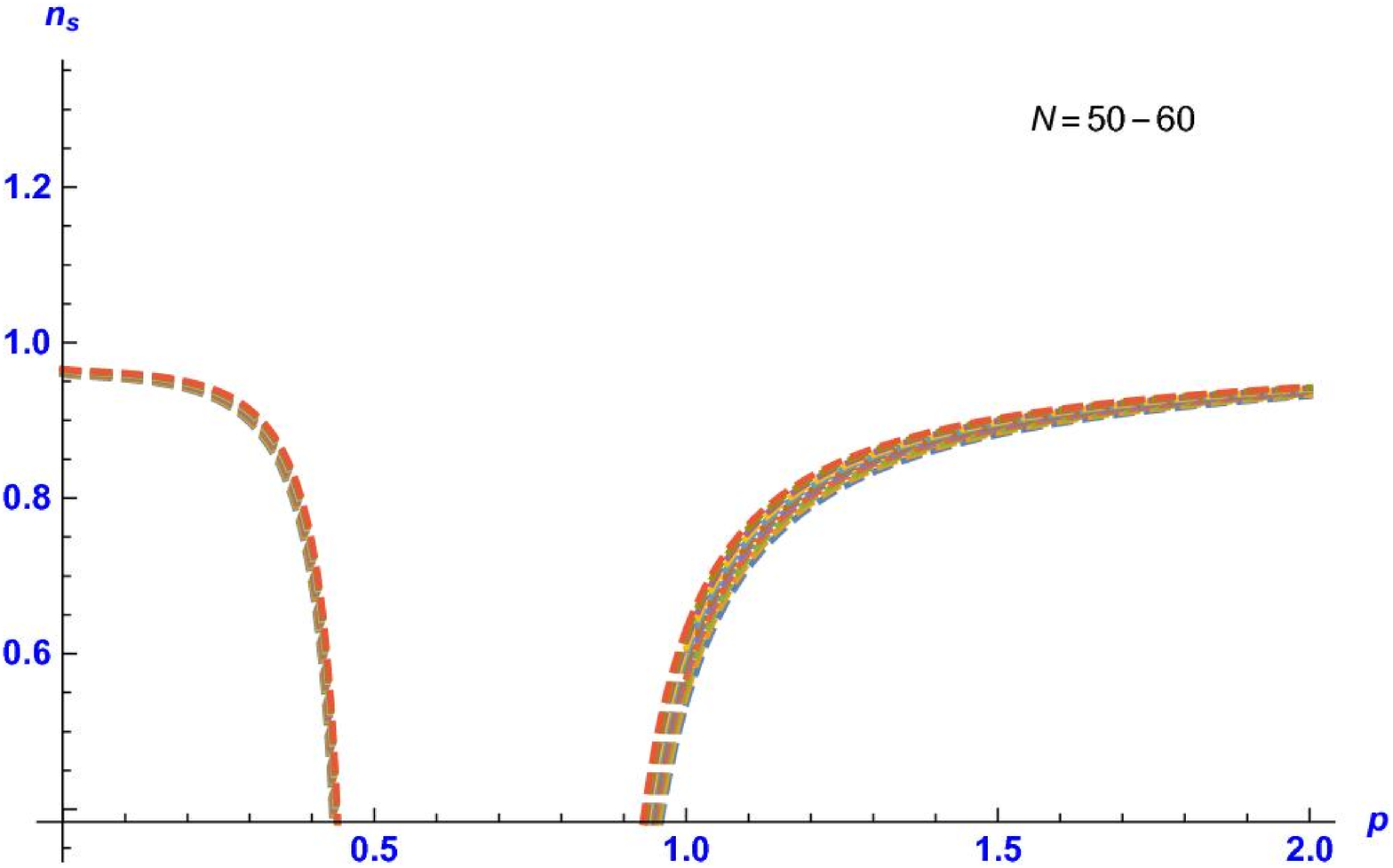}
}
 \caption{The spectral index $n_{s}$ in terms of $p$ for number of e-folds
  ($N=50-60$).}
 \label{fig7}
 \end{center}
 \end{figure}
\begin{figure}[h!]
 \begin{center}
 {
 \includegraphics[height=5cm,width=5cm]{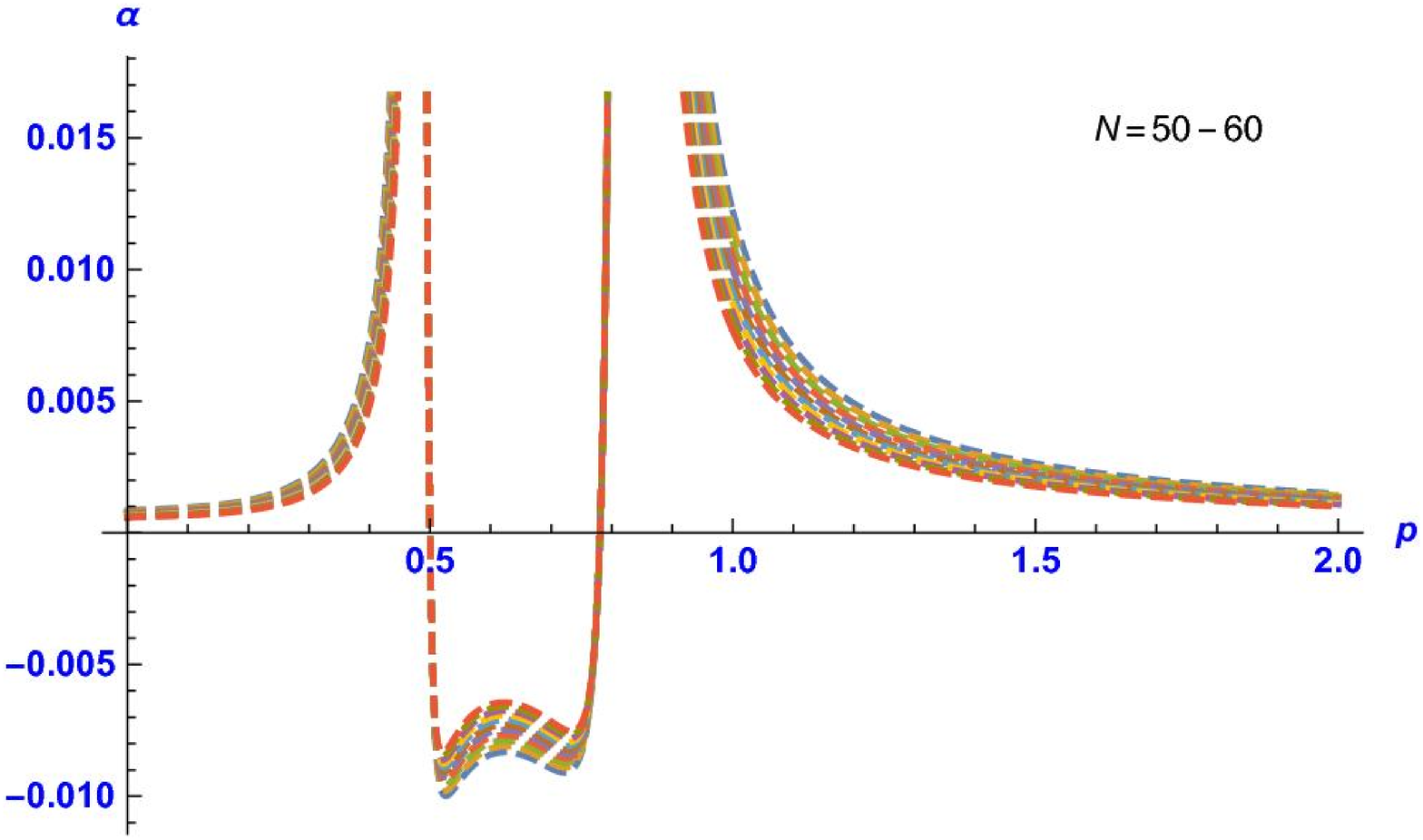}
 }
 \caption{The running spectral index $\alpha$ in terms of $p$ for number of e-folds ($N=50-60$).}
 \label{fig8}
 \end{center}
 \end{figure}
\begin{figure}[h!]
 \begin{center}
  \subfigure[] {
 \includegraphics[height=5cm,width=5cm]{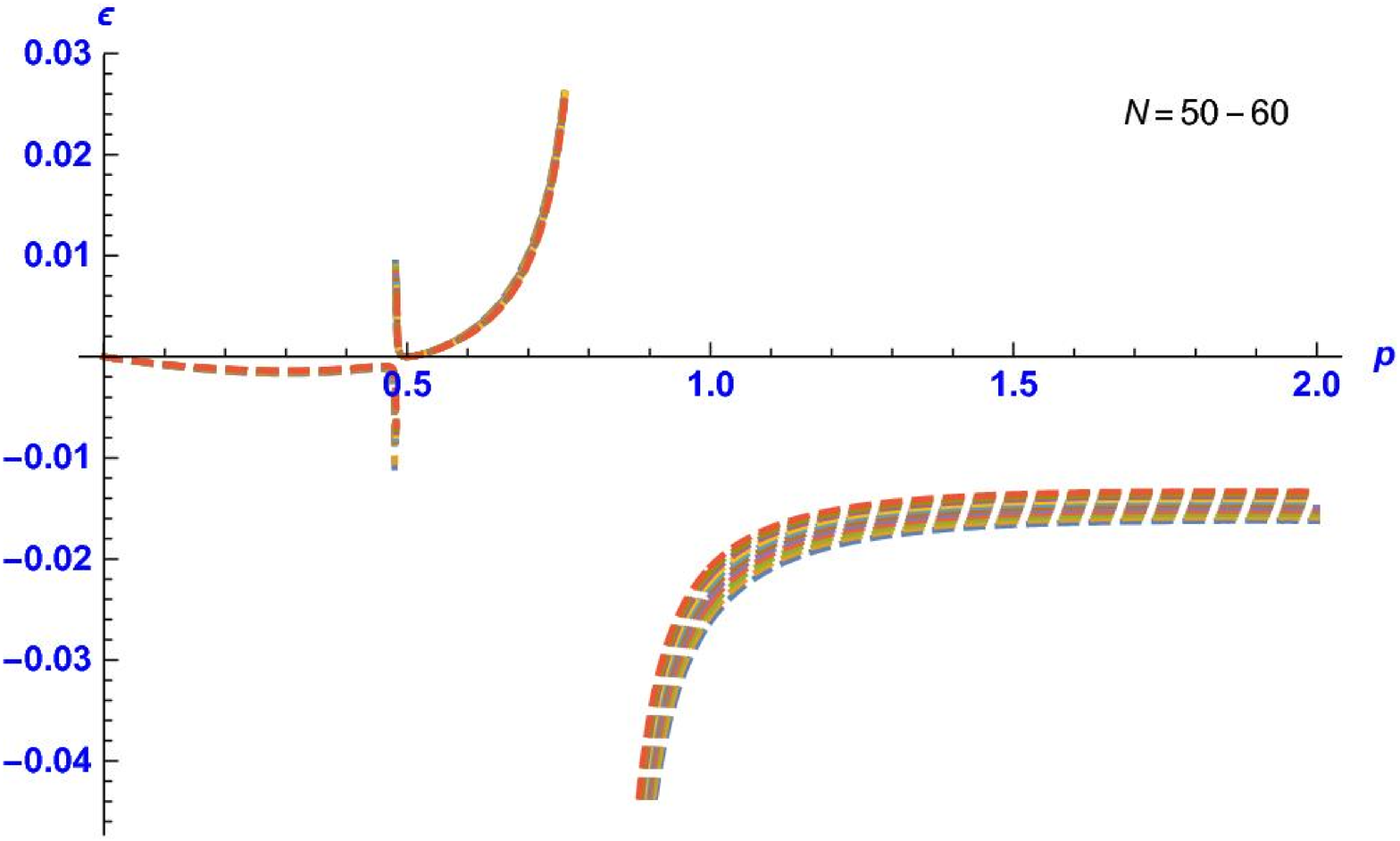}
 \label{9a}}
    \subfigure[]{
 \includegraphics[height=5cm,width=5cm]{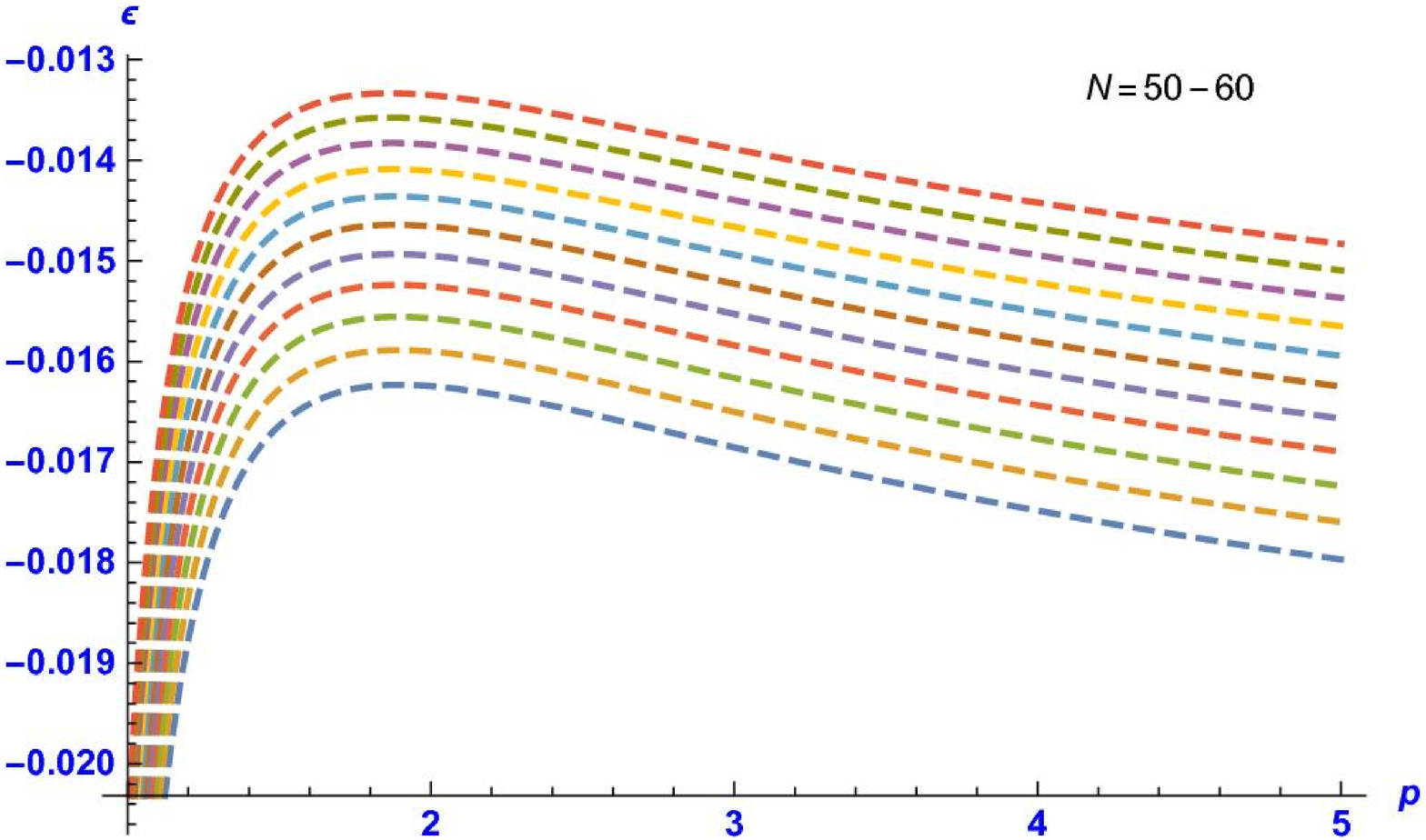}
 \label{9b}}
  \caption{\small{The slow-roll parameter $\epsilon$ in terms of $p$ for $N=50-60$ in (a) and (b) is close up of (a). }}
 \label{fig9}
 \end{center}
 \end{figure}
\begin{figure}[h!]
 \begin{center}
   \subfigure[]{
 \includegraphics[height=4cm,width=5cm]{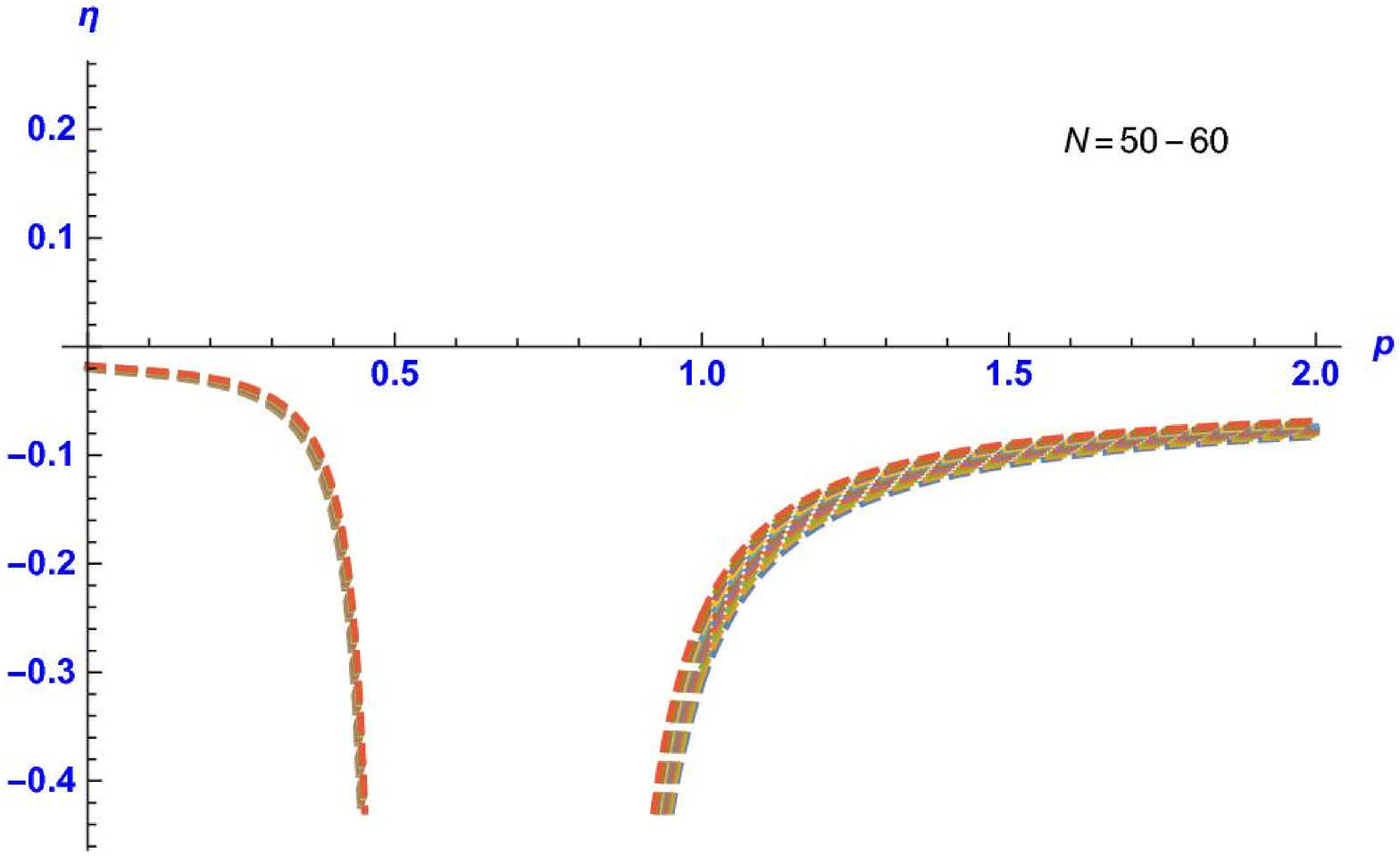}
 \label{10a}}
   \subfigure[]{
 \includegraphics[height=4cm,width=5cm]{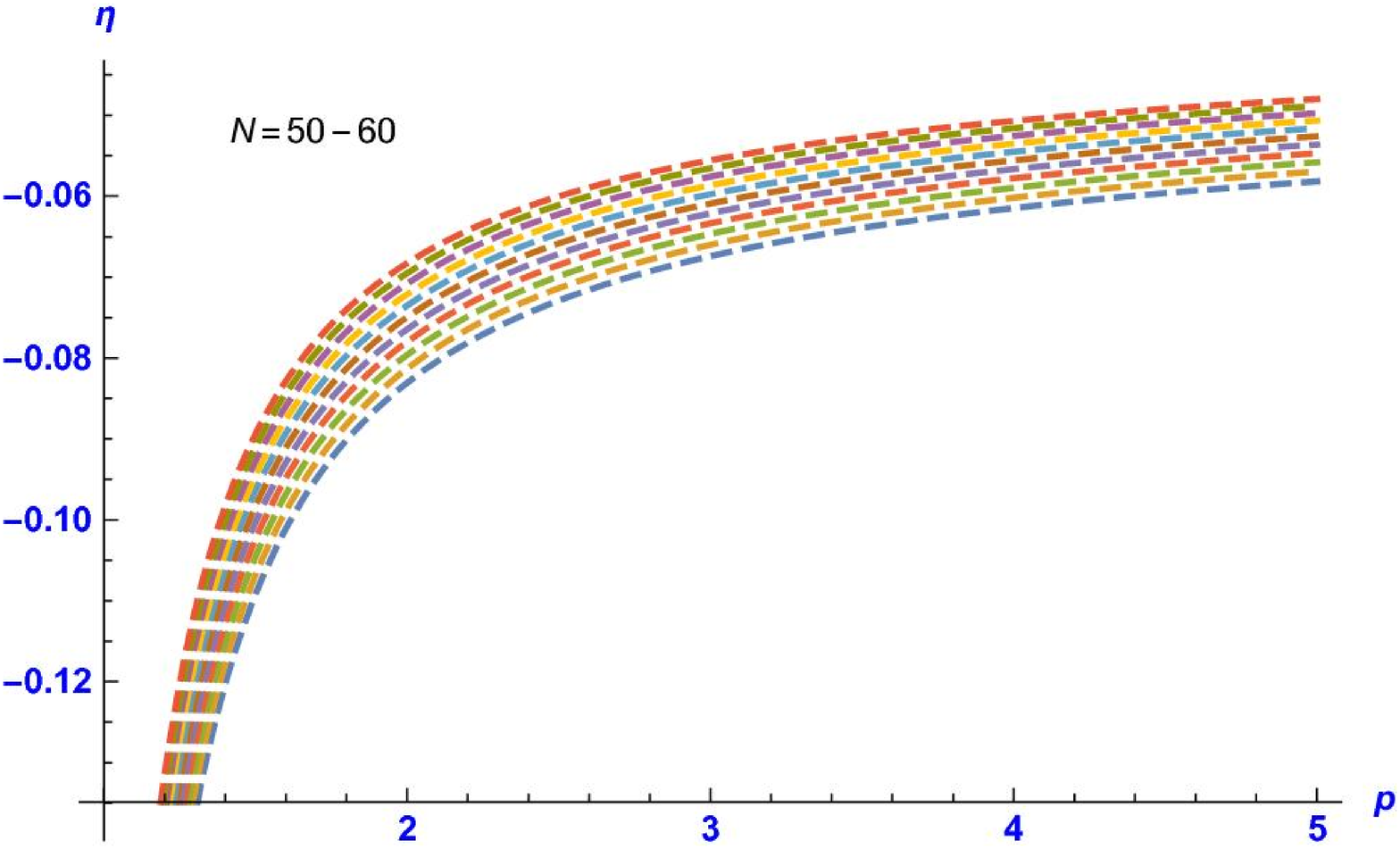}
 \label{10b}}
  \caption{\small{We plot  the slow-roll parameter  $\eta$ in terms of  $p$ for $N=60$ in (a) and  (b) is close up of (a).  }}
 \label{fig10}
 \end{center}
 \end{figure}
 \begin{figure}[h!]
 \begin{center}
  \subfigure[] {
 \includegraphics[height=4cm,width=5cm]{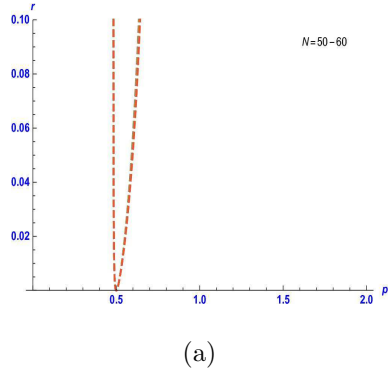}
 }
  \caption{\small{The plot of $r$ in terms of  $p$ for $N=50-60$.}}
 \label{fig11}
 \end{center}
 \end{figure}
\begin{figure}[h!]
 \begin{center}
   \subfigure[] {
 \includegraphics[height=4cm,width=5cm]{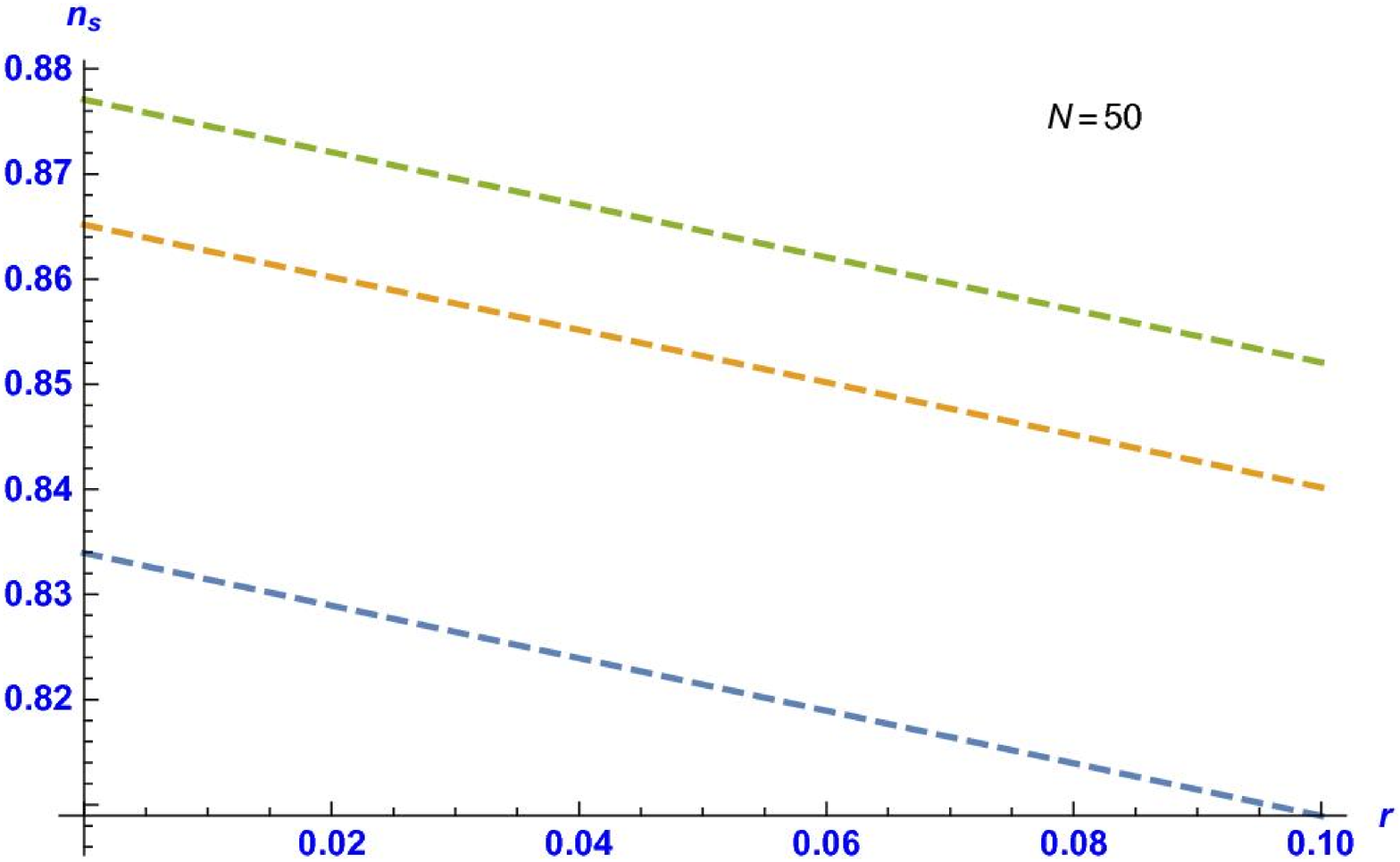}
 \label{12a}}
   \subfigure[] {
 \includegraphics[height=4cm,width=5cm]{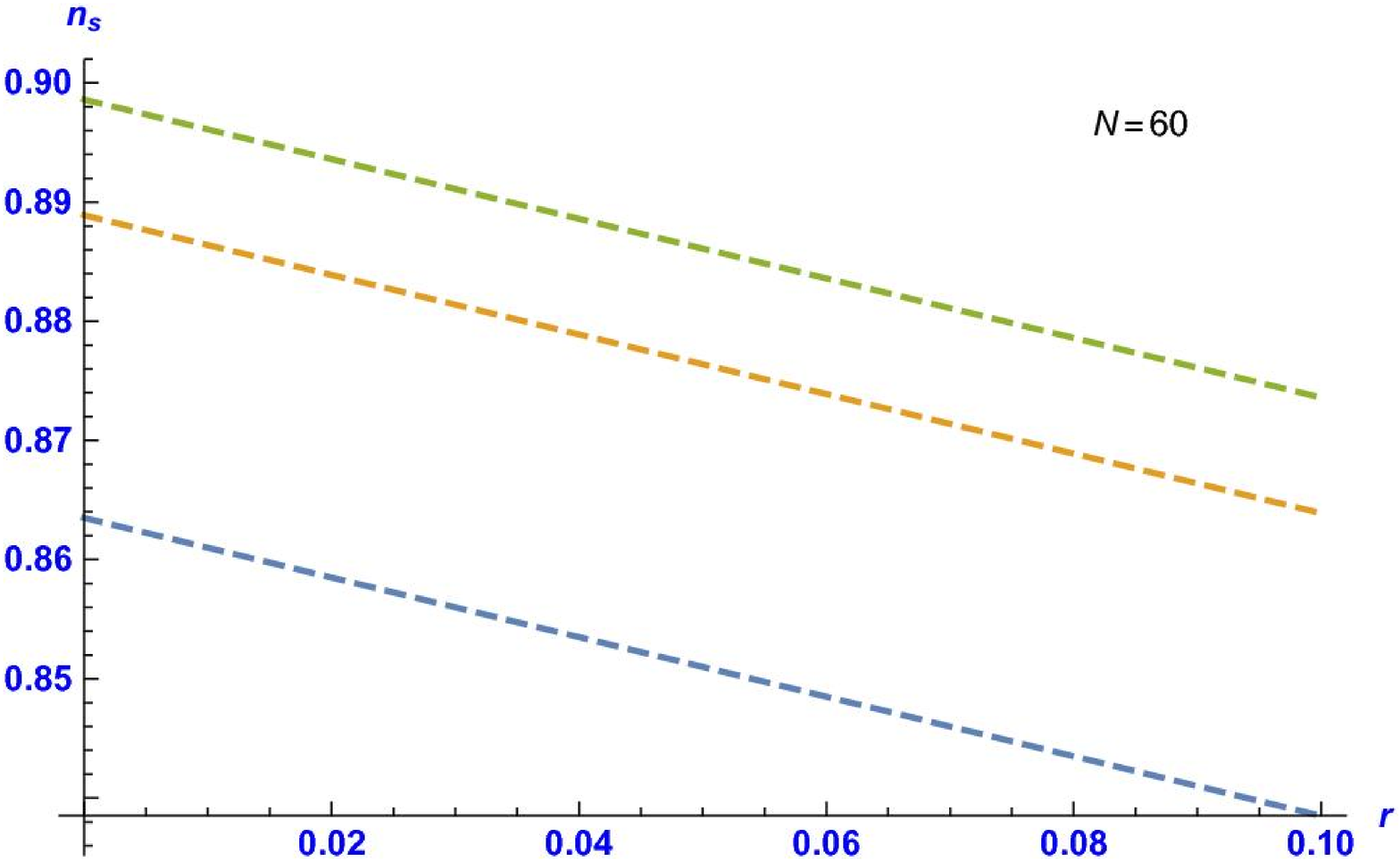}
 \label{12b}}
  \subfigure[] {
 \includegraphics[height=4cm,width=5cm]{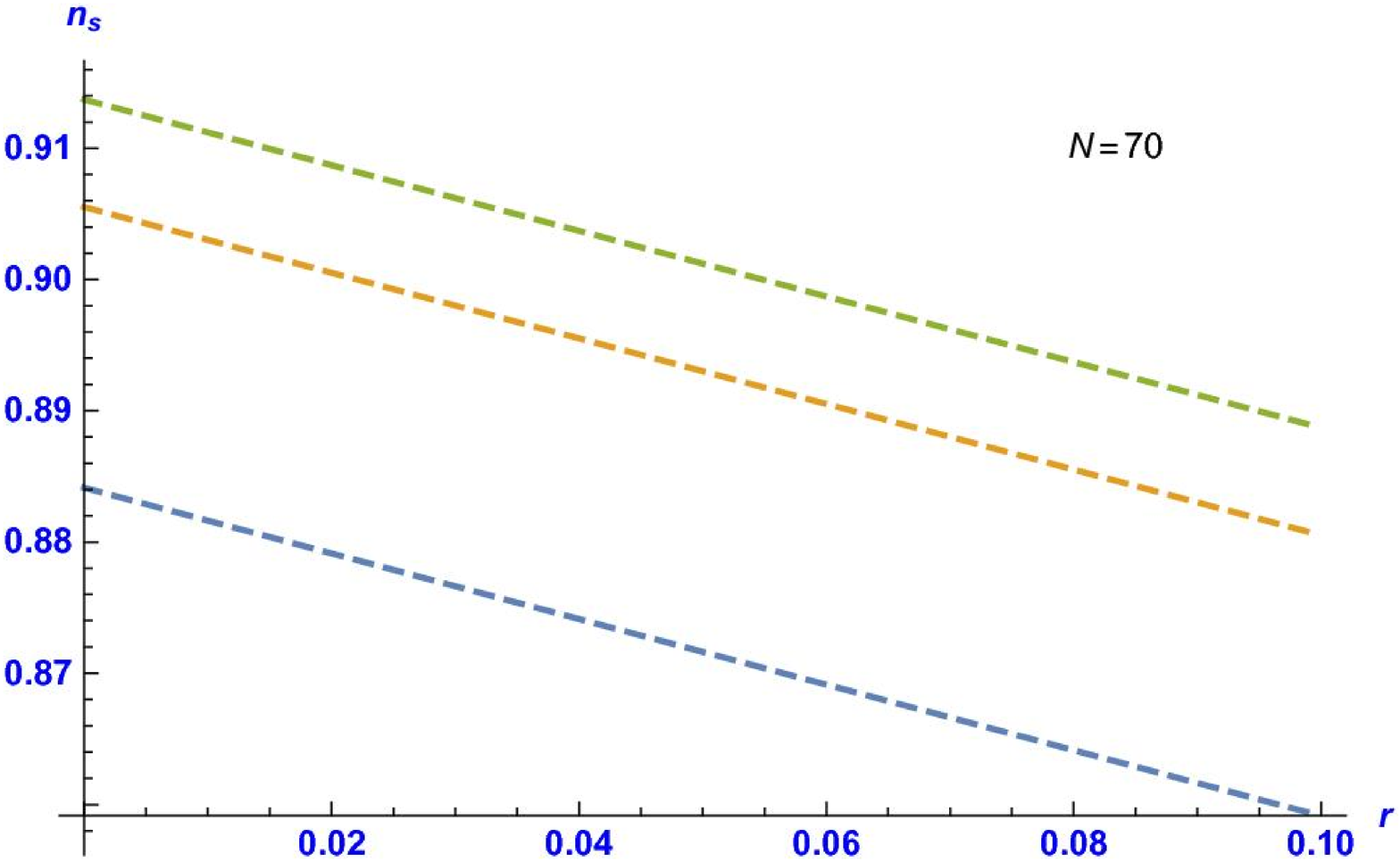}
 \label{12c}}
  \caption{\small{The plot $n_{s}$  in terms of $r$ for $N=50$, $N=60$ and $N=70$ with different value of $p$ as blue ($p=2$),  brown ($p =3$), and green ($p =4$). }}
 \label{fig12}
 \end{center}
 \end{figure}

 In general, the surprising point is the independence of these cosmological parameters to $\Lambda$. As you can see in the plots of the scalar spectrum index (Fig. \ref{fig7}), the running spectrum index (Fig. \ref{fig8}), and the slow roll parameters (Figs.  \ref{fig9} and \ref{fig10}) that the range is associated with   different values of $p$  and $N$. The range of each parameter is specified in these plots is compared to the previous plots (without contribution of $\phi_{e}$) and found that it has more minor changes. But in this case, too, for certain values, the results match with the observable data. We plot some figures to investigate each of these cosmological parameters, according to the different values of $p$, $N$, and observable data.  To be precise, we plot the $r$ in terms of  $p$ with respect to $N=50-60$ in figure (\ref{11}). The validity of this parameter is determined according to Planck's data which is $0.064$.  Also we plot $n_{s}$   for $N=50$, $N=60$ and $N=70$ and different value of $p$  in figure \ref{12}. We find  in the figure that  for   e-folds ($N = 50, 60$), the values are closer to the observable data.

\begin{figure}[h!]
 \begin{center}
   \subfigure[]{
 \includegraphics[height=6cm,width=6cm]{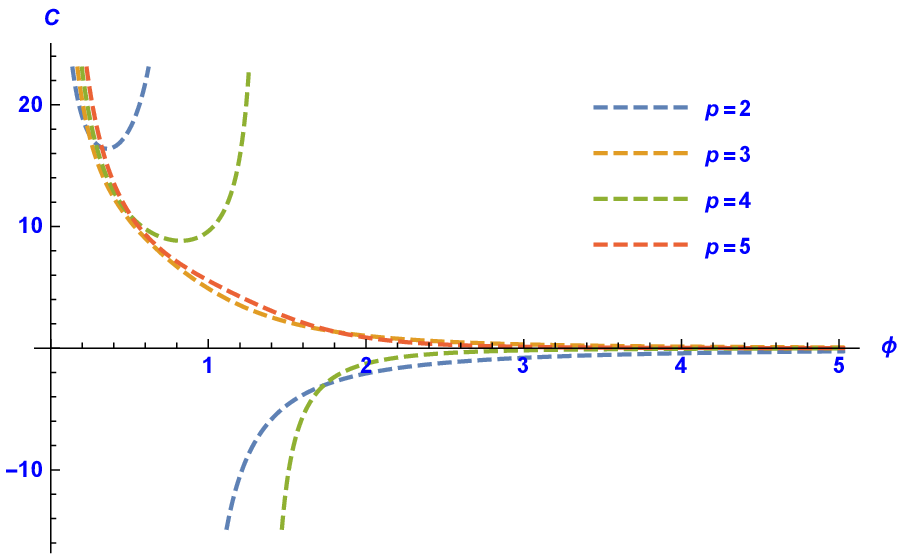}
 \label{5a}}
   \subfigure[]{
 \includegraphics[height=6cm,width=6cm]{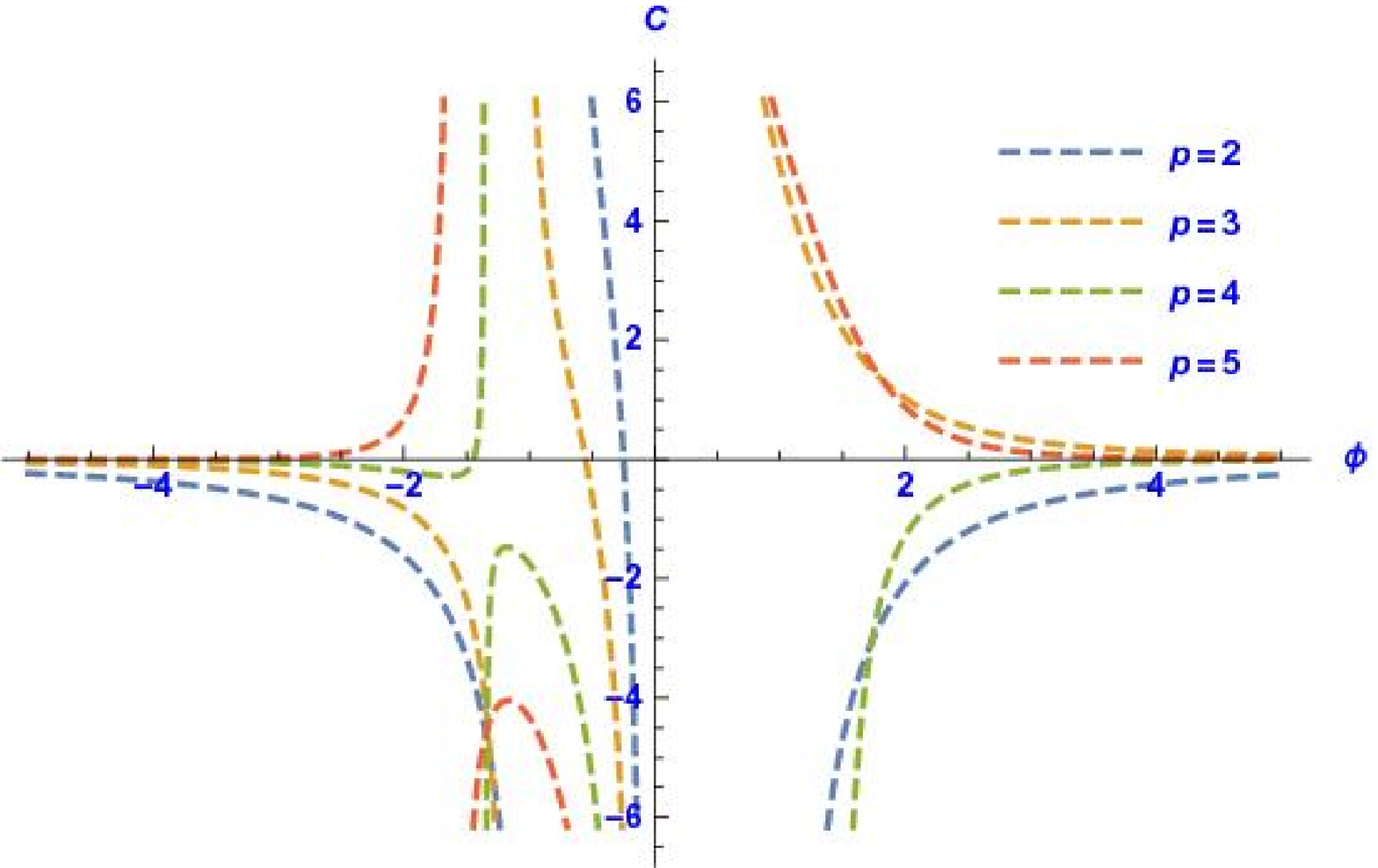}
 \label{5b}}
  \caption{\small{The plot of  $C=V'/V$ in terms of $\phi$  with respect to various values of $p$.}}
 \label{fig13}
 \end{center}
 \end{figure}

\begin{figure}[h!]
 \begin{center}
   \subfigure[]{
 \includegraphics[height=6cm,width=6cm]{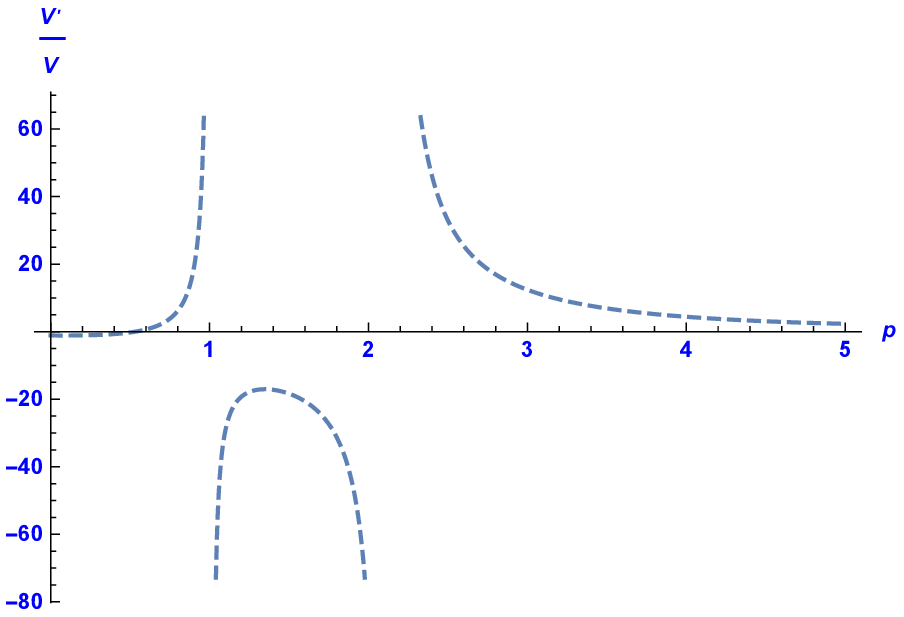}
 \label{6a}}
   \subfigure[]{
 \includegraphics[height=6cm,width=6cm]{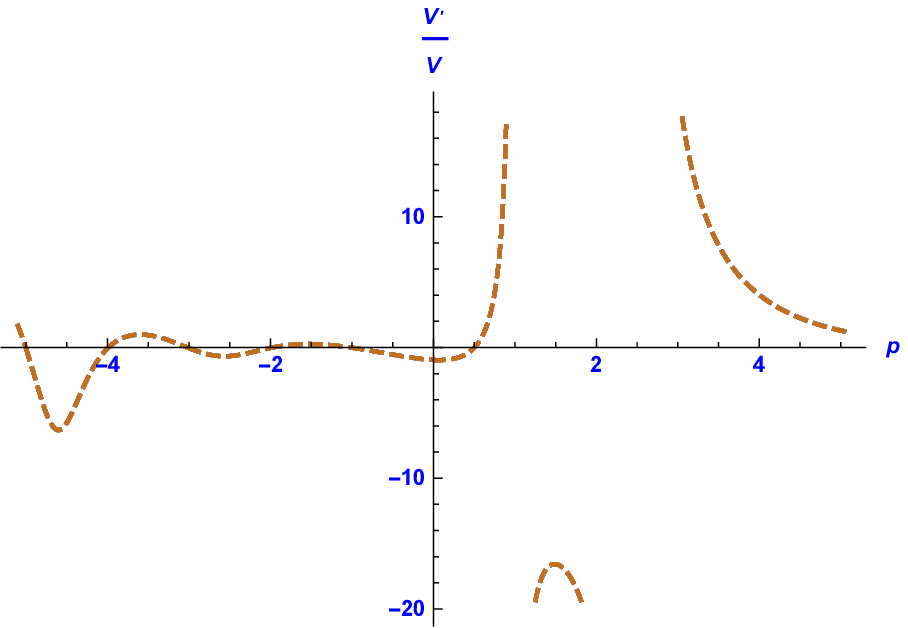}
 \label{6b}}
  \caption{\small{The plot of  $V'/V$ in terms of $p$  with respect to $N=60$ in fig (6a), and $N=50-60$  in fig (6b).}}
 \label{fig14}
 \end{center}
 \end{figure}

 Finally, we can specify the values and range associated with each of the $ \phi $ and $ \Lambda $ parameters by using the calculated values and observable data. In fact, in this paper, we examined the modified gravitational model on the brane and its compatibility with the swampland criteria. The interesting thing is that the cosmological parameters were independent of the $ \Lambda $ parameter. The present studies  can also be extended for other inflation models and pay attention to the existing commonalities. An important point to note is whether the model studied in this paper falls within the range of swampland conjectures or clearly how much of the parameter space of the modified gravitational model falls within the range of the swampland conjecture.
What role will each of the important cosmological parameters play in this model?
Is this compatibility in line with the latest cosmological data, or can it be examined more broadly?
Therefore, as shown in (Figs.  \ref{fig13} and \ref{fig14}), the parametric range allowed for the modified gravitational model in terms of the scalar field $ \phi $ and the constant parameters mentioned in the text, especially the parameter (p) and with respect to the condition $V/\Lambda >> 1$ is specified. As shown in (Fig.  \ref{fig13}), we have a plot of $ V '/ V $ in terms of $ \phi $ with respect to various values of $p$, and as you can see in a specific region, the relation $V' /V$ on these plots is order one or greater, as the swampland conjecture would demand. Of course, we tried to determine the range for the negative values of the scalar field.
Of course, unauthorized areas are clearly identified in the figures, and also the constant parameters mentioned play their role well.
Similarly, according to equations (\ref{18}) and (\ref{25}), the changes in swampland conjecture are shown according to the constant parameter $p$ for different a number of e-folds N in (Fig.  \ref{fig14}). The compatibility and alignment of these swampland conjectures with each of the cosmological parameters are also determined. Thus, the coordinated gravitational model and swampland conjectures in a particular parametric space can be used.
Of course, deeper challenges can also be posed to swampland conjectures for such models, where the main focus can be on classifying a new type of inflation model according to swampland criteria, provided that swampland conjecture can be used in the future for Adapting to existing theories and solving many problems in modern cosmology.
 It can now be seen as a growing emerging theory.

\begin{figure}[h!]
 \begin{center}
   \subfigure[]{
 \includegraphics[height=6cm,width=6cm]{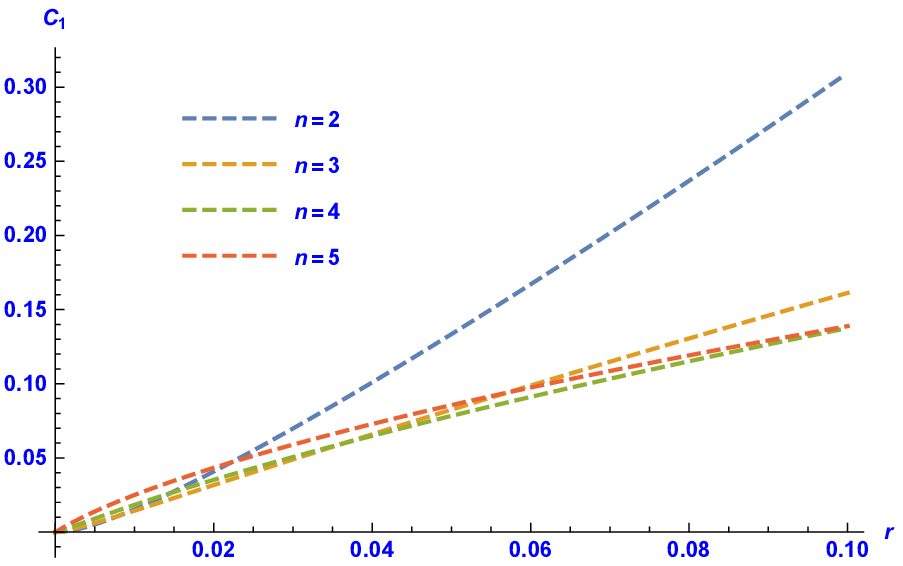}
 \label{7a}}
   \subfigure[]{
 \includegraphics[height=6cm,width=6cm]{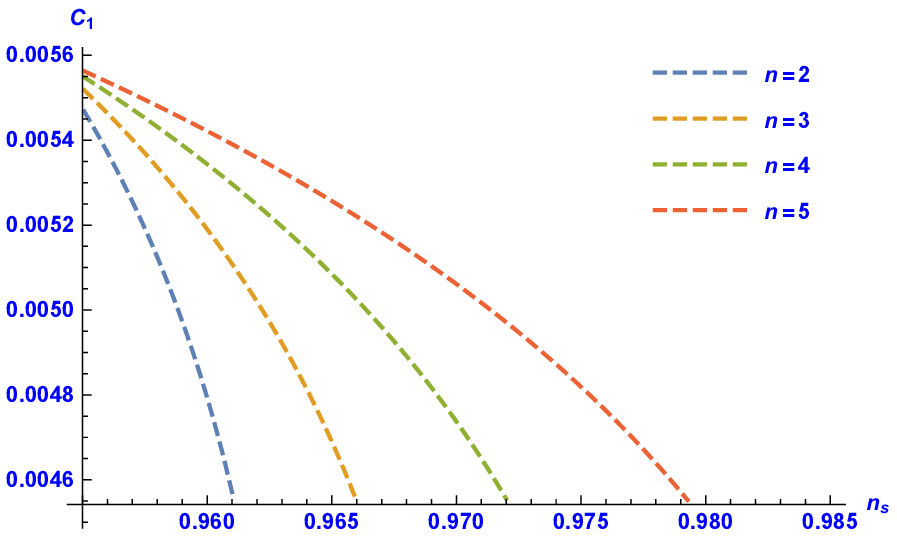}
 \label{7b}}
  \caption{\small{The plot of  $C=V'/V$ in terms of $r$  and $n_{s}$ with respect to various values of $p$.}}
 \label{fig15}
 \end{center}
 \end{figure}

In addition to the items mentioned in the text of the article, swampland conjectures can be challenged in other ways, and these conjectures can also be examined about cosmological parameters.
 Among these, two very important cosmological parameters, i.e., the scalar spectrum index $n_{s}$ and tensor-to-scalar ratio $r$, can be examined according to the swampland conjectures. The allowable range of each of these two cosmological parameters can be determined in relation to the swampland conjecture components.
That is, in fact, using equations (\ref{18}), (\ref{32}), (\ref{35}), and (\ref{36}), we challenged the first conjecture of the swampland in terms of these two cosmological parameters, namely $n_{s}$  and $r$.
As you can see in Fig. \ref{fig15}, the allowable range for each of these cosmological parameters and the first component of the swampland are determined according to the constant parameter $p$.
As it is clear in the literature, the first and second components of the swampland are always a positive value and of unit order.
In this figure, although these two parameters are within the allowable range of the swampland components, their changes are also specified for the changes of the constant parameter $p$.
The allowable range for the second component can be calculated in the same way, and it can be determined what virtual range can be specified for different fixed values.
In fact, by challenging the first swampland conjecture, we found that this satisfaction and acceptance in relation to the model mentioned is clearly indicated in Fig. \ref{fig15}.

\section{Conclusions}\label{sec4}
Recently, various inflation models have been studied by researchers from different perspectives and conditions such as slow-roll, constant-roll, ultra-slow-roll,  and weak gravity conjecture in order to introduce a model for expanding the universe. In this paper, we investigated a new condition for inflation models by introducing a modified ($R+\gamma R^{p}$) gravitational model. In this paper, we considered the special and interesting case of $\gamma=1$, so it is interesting also to consider other values of $\gamma$ in future works. As our studies are based on a modified  $f(R)$ gravitational model on the brane,  we encountered modified cosmological parameters. So, we first introduced these modified cosmological parameters such as spectral index, a number of e-folds and etc. Then we applied these conditions to the modified $f(R)$ gravitational model in order to adapt to the swampland criteria. Finally, we determined the range of each of these parameters by plotting some figures and with respect to observable data such as Planck 2018. A very interesting and exciting point we observed is that  the series of cosmological parameters throughout the calculations do not depend on $\Lambda$.

\end{document}